\documentclass[twoside,reqno,11pt]{amsart}
\usepackage{a4wide}
\usepackage{amsmath}
\usepackage{amsfonts}
\usepackage{amssymb}
\usepackage{times}
\usepackage[usenames,dvipsnames]{pstricks}
\usepackage[arrow, matrix, curve]{xy}
\usepackage{pstricks,pst-node,pst-text,pst-3d}
\usepackage{pst-grad}
\usepackage{graphicx}
\DeclareGraphicsExtensions{.ps}

\psset{framearc=0.0,framesep=0.0truecm,nodesep=3pt,unit=0.75truecm,%
arrowsize=4pt 2,arrowinset=0.4}
\def\pstlw{0.8pt}
\newcounter{mycnt}
\def\themycnt{\thesection.\arabic{mycnt}}
\def\mybenv#1{\refstepcounter{mycnt}%
       \vskip 3pt\noindent{\bf #1~~\themycnt}:~}
\def\myeenv{\hfill\rule{1ex}{1ex}\vskip 3pt}
\def\qed{\hfill$\Box$}
\def\nn{\nonumber \\}
\def\Id{\text{I\!d}}

\def\openR{\mathbb{R}}
\def\openC{\mathbb{C}}

\def\openZ{\mathbb{Z}}

\def\CO{\Delta}

\def\ot{\otimes}

\def\antip{{\sf S}}

\def\la{\langle}
\def\ra{\rangle}

\def\!{\kern -0.15ex}
\def\antip{\textsf{S}}

\def\grpGL{\textsf{GL}}
\def\grpH{\textsf{H}}

\def\CharGL{{\sf Char-GL }}
\def\CharO{{\sf Char-O }}
\def\CharSp{{\sf Char-Sp }}
\def\CharHpi{\textsf{Char-H$_\pi$}}
\def\comp{\raisebox{0.2 ex}{${\scriptstyle \circ}$}}
\def\piprod{\raisebox{0.2 ex}{${\scriptstyle \odot}$}\kern .2ex}
\def\bcap{{\texttt b}}
\def\dcup{{\texttt d}}
\def\bpcap{\overline{\texttt b}{}_\pi}
\def\dpcup{\overline{\texttt d}{}_\pi}
\begin{document}

\title[Ribbon Hopf algebras]
{Ribbon Hopf algebras from group character rings\footnote{T\lowercase{his
research was supported through the programme} ``R\lowercase{esearch in}
P\lowercase{airs}'' \lowercase{by the} M\lowercase{athematisches}
F\lowercase{or\-schungsinstitut} O\lowercase{berwolfach in} 2010.}} 
\author{Bertfried Fauser}
\address{%
School of Computer Science\\
The University of Birmingham\\
Edgbaston-Birmingham, W. Midlands, B15 2TT, England}
\email{fauser.b@cs.bham.ac.uk}
\author{Peter D. Jarvis}
\address{%
School of Mathematics and Physics, University of Tasmania, 
Private Bag 37, GPO, Hobart Tas 7001, Australia}
\email{Peter.Jarvis@utas.edu.au}
\author{Ronald C. King}
\address{%
School of Mathematics, University of Southampton,
Southampton SO17 1BJ, England}
\email{R.C.King@soton.ac.uk}



\date{June 12, 2011}

\begin{abstract}
We study the diagram alphabet of knot moves associated with the character
rings of certain matrix groups. The primary object is the Hopf algebra
{\sf Char-GL} of characters of the finite dimensional polynomial
representations of the complex group $GL(n)$ in the inductive limit,
realised as the ring of symmetric functions $\Lambda(X)$ on countably
many variables $X =  \{ x_1,x_2,\cdots \}$. Isomorphic as spaces
are the character rings {\sf Char-O} and {\sf Char-Sp} of the classical matrix
subgroups of $GL(n)$, the orthogonal and symplectic groups. We also analyse
the formal character rings {\sf Char-H$_\pi$} of algebraic subgroups of
$GL(n)$, comprised of matrix transformations leaving invariant a fixed but
arbitrary tensor of Young symmetry type $\pi$, which have been introduced
in~\cite{fauser:jarvis:king:wybourne:2005a} (these include the orthogonal and
symplectic groups as special cases). The set of tangle diagrams encoding
manipulations of the group and subgroup characters has many elements deriving
from products, coproducts, units and counits as well as different types of
branching operators. From these elements we assemble for each $\pi$ a crossing
tangle which satisfies the braid relation and which is nontrivial, in spite of
the commutative and co-commutative setting. We identify structural elements and
verify the axioms to establish that each {\sf Char-H$_\pi$} ring is a ribbon
Hopf algebra. The corresponding knot invariant operators are rather weak,
giving merely a measure of the writhe.
\end{abstract}

\maketitle

\section{Introduction}
\label{sec:Intro}
In this article we develop a new approach to realisations of the braid group,
and indirectly to knot and link invariants.  In standard diagrammatic
treatments, lines are decorated with linear spaces, typically modules of a
suitably deformed algebra, and crossings encode the action of appropriate
operators or $R$-matrices representing braid generators. By contrast, our lines
in tangle diagrams are decorated with group characters, which do not entail any
deformation of the groups. Instead, the `deformation' which gives rise to
nontrivial braidings, is combinatorially generated at the level of the product
rule for the group characters themselves.

By way of introduction we now elaborate on these statements in order to motivate
and explain the strategy underlying our work, and to clarify our claims. We
shall not dwell on the above-mentioned deformed algebra approach, which has been
well-studied in the literature and is covered in many standard
texts~\cite{turaev:1994a,kauffman:1991a,ohtsuki:2002a}. 
However, we shall make central use of graphical calculus,
and we pause here to explain briefly the origins of its use in knot theory. Then
we indicate its extension beyond the knot alphabet, to tangle elements arising
from the Hopf structure of the character rings, which we shall need for
character manipulations. Indeed, the paper as a whole can be seen as a
description of how the extended tangle alphabet can be used to assemble a
useable graphical calculus for knots and links. The present introduction can
also be read as an informal outline of the paper. A formal resum\'{e} is also
provided at the end of this section.

Let $S_1$ be a circle (or let $S_1^{\times k}$ be a direct product of
$k$-circles in case of a link on $k$ strands). A \emph{knot} is a tame nonsingular
embedding of the circle into $\openR^3$. This means that the embedded curve is
smooth (no cusps with bounded extrinsic curvature) and does not have
self-crossings, hence does not have singular points. By considering projections
$\pi_T:K\rightarrow T$ of the knot into the plane, we obtain \emph{tangles}
which keep the over and under information. It is known that there is an
isomorphism from knot isotopy classes to isotopy classes of tangles
\cite{ohtsuki:2002a}, preserving all topological information about the knot.
Finally, we fix a particular projection, as a representative of the tangle
isotopy. Now by the fundamental Alexander theorem, every knot can be deformed
via an isotopy into a closed braid. The closed braids obtained by
projection index the isotopy classes of knots, and the braid group
representations become central in studying knots and their invariants, as illustrated schematically in 
(eqn: \ref{fig:3dProjection}).
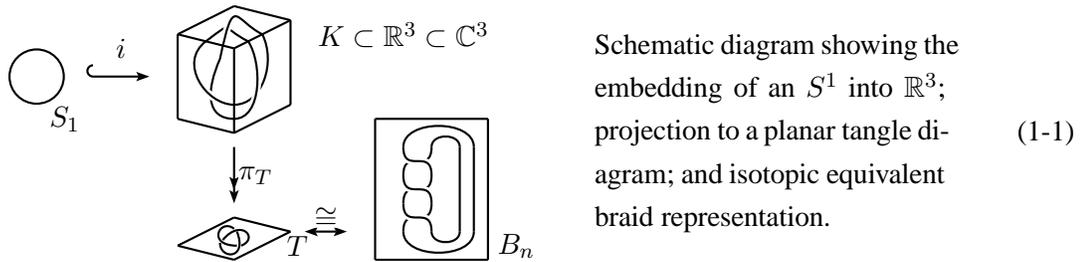
\begin{align}\label{fig:3dProjection}
\vcenter{\hsize=0.4\textwidth\mbox{\pspicture(0,0)(5,4.5)
\psset{linewidth=.8pt,xunit=0.5,yunit=0.5,runit=0.5}
\psset{arrowsize=2pt 2,arrowinset=0.2}
%
%
\pscircle[linewidth=.8pt](1,6.5){1.0}
\rput(2,5){\psframebox*[framesep=0pt]{$S_1$}}
\psarc[linewidth=\pstlw](3,6.7){0.2}{90}{270}
\psline[linewidth=\pstlw]{->}(3,6.5)(5,6.5)
\rput(4,7.5){\psframebox*[framesep=0pt]{$i$}}
%
%
\psbezier(8.5,7.75)(7.5,8.25)(8,8.5)(7,7.75)
\psbezier[border=1pt,bordercolor=white](9,8)(8,8.5)(8,9.5)(7.5,7.125)
\psbezier(7.5,7.125)(7,6)(7,5)(8,5)
\psbezier[border=1pt,bordercolor=white](7,7.75)(6,7)(7,5.5)(8,5.5)
\psbezier(8,5.5)(9,5.5)(9.5,7)(9,8)
\psbezier[border=1pt,bordercolor=white](8.5,7.75)(9,7)(9,5)(8,5)
%
%
\psline{-}(6,5)(6,8)
\psline{-}(6,5)(8,4.5)
\psline{-}(8,4.5)(10,5.5)
\psline{-}(8,4.5)(8,7.5)
\psline{-}(10,5.5)(10,8.5)
\psline{-}(6,8)(8,7.5)
\psline{-}(8,7.5)(10,8.5)
\psline{-}(6,8)(8,9)
\psline{-}(8,9)(10,8.5)
\rput(14,8){\psframebox*[framesep=0pt]{$K\subset\openR^3\subset\openC^3$}}
\psline{->>}(8,4)(8,2)
\rput(8.75,3){\psframebox*[framesep=0pt]{$\pi_{T}$}}
%
%
\psline{-}(6,0.5)(8,0)
\psline{-}(8,0)(10,1)
\psline{-}(10,1)(8,1.5)
\psline{-}(8,1.5)(6,0.5)
%
%
\psbezier(8,1.1)(7.5,1.5)(7,0.5)(8.1,0.5)
\psbezier(8.25,0.5)(8.9,0.75)(8.25,1.5)(7.6,0.7)
\psbezier(7.5,0.6)(7.5,0)(8.5,0.0)(8.1,1)
\rput(10.25,0.5){\psframebox*[framesep=0pt]{$T$}}
\psline{<->}(10.5,1)(12,1)
\rput(11.25,1.5){\psframebox*[framesep=0pt]{$\cong$}}
%
%
\psline{-}(13,0)(17,0)
\psline{-}(17,0)(17,5)
\psline{-}(17,5)(13,5)
\psline{-}(13,5)(13,0)
%
%
\psbezier(15,4)(15,3)(14,4)(14,3)
\psbezier[border=1pt,bordercolor=white](14,4)(14,3)(15,4)(15,3)
\psbezier(15,3)(15,2)(14,3)(14,2)
\psbezier[border=1pt,bordercolor=white](14,3)(14,2)(15,3)(15,2)
\psbezier(15,2)(15,1)(14,2)(14,1)
\psbezier[border=1pt,bordercolor=white](14,2)(14,1)(15,2)(15,1)
\psbezier(14,4)(14,5)(16.5,5)(16.5,4)
\psbezier(15,4)(15,4.5)(16,4.5)(16,4)
\psbezier(14,1)(14,0)(16.5,0)(16.5,1)
\psbezier(15,1)(15,0.5)(16,0.5)(16,1)
\psline{-}(16,1)(16,4)
\psline{-}(16.5,1)(16.5,4)
\rput(18,0.5){\psframebox*[framesep=0pt]{$B_n$}}
\endpspicture 
}}
&&\hskip6ex
\vcenter{\hsize=0.3\textwidth\noindent Schematic diagram showing the embedding
of an $S^1$ into $\mathbb{R}^3$; projection to a planar tangle diagram; and
isotopic equivalent braid representation.}
\end{align}
The `diagrammar' of knot moves arises from the way in which \emph{sliced
tangles} are made from the tangle diagrams. For tame knots, it is possible 
under isotopy to
distribute horizontal cuts such that at most one non-trivial, non-identity
operation occupies each horizontal strip or slice. Thus, abstractly, knot or
link tangles are generated using a realization of a knot \emph{alphabet} of
basic letters\footnote{We read tangle diagrams from top to bottom using the
`pessimistic arrow of time' \cite{oziewicz:1999}, also our crossing is
left-handed, so we are left-handed psessimists. Some literature features
the right-handed optimist reading, so care is needed comparing images.}:
\begin{align}\label{eq:KnotAlphabet}
\textrm{knot-alphabet~} &=\,\,\Big\{
\vcenter{\hsize=0.45\textwidth\input{tangles/knotAlphabet.tex}}
\Big\vert\textrm{~~relations~~}\Big\},
\end{align}
where the `relations' represent equivalence between certain words in this
alphabet, implementing the isotopy equivalence of different knot projections
represented by these tangles. The formal use of these elements in defining knot
invariants will be dealt with in the body of the
paper~(\S~\ref{sec:KnotsnLinks}).

In order to motivate the extended diagrammatic tangle alphabet which we require,
and its relation to the knot alphabet, we briefly introduce the mathematical
setting in which we work. Full details are elaborated in \S
\ref{subsec:PiNewellLittlewood}, \S \ref{sec:KnotsnLinks}, and \S
\ref{subsec:KnotAlphabet} below.

Consider the ring of characters of finite-dimensional, polynomial irreducible
representations of the complex matrix group $GL(n)$. It is well known that in
the inductive limit $n  \rightarrow \infty $, this ring {\sf Char-GL} is isomorphic
to the ring $\Lambda(X)$ of symmetric polynomials in countably many variables $X
= \{ x_1,x_2,x_3,\cdots \} $ (see \S \ref{subsec:SymmX} for notation). As will
be explained in detail in \S \ref{subsec:SymmX}, {\sf Char-GL} is a Hopf
algebra. Lines in tangle diagrams are labelled by elements of the algebra (for
example one may use the famous Schur- or $S$-functions as a basis). Both the
algebra (pointwise multiplication) and bialgebra (comultiplication) structures
require specific relations amongst combinations of diagrammatic elements.
Including certain other types of canonical structural elements, the tangle
alphabet then contains the following (possibly oriented) types of elements:
\begin{align}\label{eq:GLalphabet}
\textrm{{\sf GL}-alphabet} &= \,\left\{\!
\vcenter{\hsize=0.6\textwidth\input{tangles/GLAlphabet.tex}}
\Big\vert\textrm{~~relations~~} \right\}, \nn
&=\,\left\{
\hskip2ex\Id;
\hskip1ex\eta;
\hskip0.6ex\epsilon;
\hskip1.2ex \Delta;
\hskip1.8ex m;
\hskip1.7ex\antip;
\hskip1.3ex C[X;Y];
\hskip0.1ex\la-\mid-\ra;
\hskip0.5ex c_{U,V};
\hskip1.5ex c^{-1}_{U,V}
\hskip2.6ex\Big\vert\textrm{~~relations~~}\right\}
\end{align}
where the `relations' code for associativity and units for the products
and other structural compatibilities demanded by the Hopf algebra axioms.
Some of these symbols will need to acquire additional attributes such
as orientation; their full descriptions will of course be developed in
the body of the paper.

Beyond the ambient ring {\sf Char-GL} of symmetric functions we also require the
character rings of certain algebraic matrix subgroups $H_\pi(n)$ of $GL(n)$,
defined as complex matrices preserving a fixed complex tensor of Young symmetry
type $\pi$
\cite{fauser:jarvis:king:wybourne:2005a,fauser:jarvis:king:2005a,fauser:jarvis:king:2007c}.
The rings {\sf Char-H}$_\pi$ are isomorphic to {\sf Char-GL} as linear spaces,
and have diagrammatic elements derived
from the {\sf GL}-alphabet but subject to their own relations. Using these
elements as building blocks, we can assemble composite 2-2 tangles, or
crossings, which are nontrivial, and which satisfy the braid relations
(necessarily so, as a result of the Hopf structure). In fact by identifying
certain universal algebraic elements, and appropriately completing the alphabet
of knot moves, we prove in \S \ref{subsec:KnotAlphabet} the main theorem of this
paper (Theorem \ref{thm:MainTheorem}), that each {\sf Char-H}$_\pi$ ring has the
structure of a ribbon Hopf algebra \cite{turaev:1994a,kassel:1995a} and so can
be used to represent knots.

The emergence of knot and link representations in a commutative, co-commutative
setting is a striking result from the viewpoint of standard approaches using
deformed algebras and braid operators to represent braid generators. As
mentioned, we deal with standard matrix groups. However, from a Hopf algebraic
perspective, there is indeed a deformation \cite{sweedler:1968a,sweedler:1969a},
because of the changed character product rule  (the $\pi$-Newell-Littlewood rule
\cite{fauser:jarvis:king:wybourne:2005a}, which differs from pointwise
multiplication), enjoyed by symmetric functions as elements of {\sf
Char-H}$_\pi$ rather than as elements of {\sf Char-GL}. This deformation is
combinatorially labelled by each distinct symmetry type represented by the
partition $\pi$, rather than being labelled by a parameter. 

This situation can be described dually in the following way. Recall that the
symmetric group ${\mathfrak S}_n$ plays the role of the centraliser algebra of
the general linear group in tensor products. However, ${\mathfrak S}_n$ is
replaced by a different algebra for matrix subgroups of the general linear group
(for example, it is replaced by the Brauer algebra for the orthogonal and
symplectic groups). Labelling knot tangle diagrams or closed braids with group
characters, the conversion of an element of the braid group ${\mathfrak B}_n$ to
an operation on characters in {\sf Char-GL} can schematically be viewed as
passing from ${\mathfrak B}_n$ to ${\mathfrak S}_n$, whereby all crossing
information is lost, with the replacement of braid moves merely by
transpositions. However, the labelling with characters in {\sf Char-H$_\pi$}
entails operations controlled by the appropriate centralizer algebra. Working in
{\sf Char-GL}, we arrive at a tangle which is a pull-back of the isomorphism
between the underlying spaces, which is no longer equivalent to the
directly-derived, degenerate tangle, and hence may retain topological
information (eqn: \ref{fig:CentralizerSchema}).
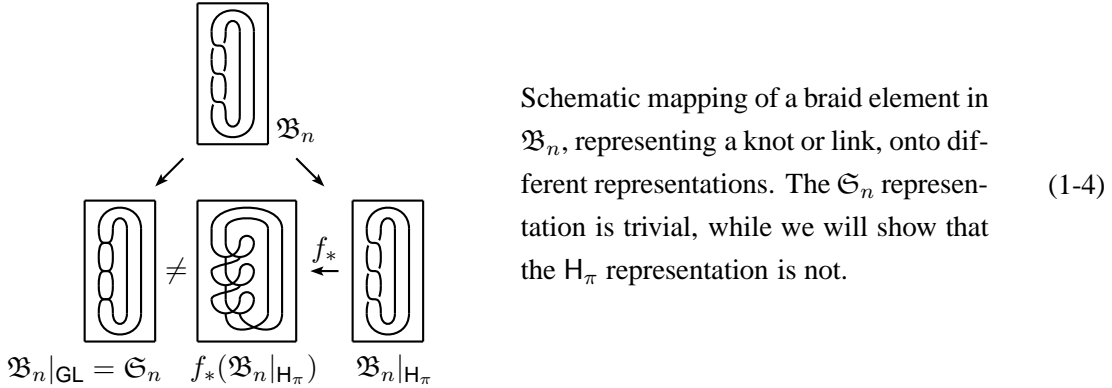
\begin{align}\label{fig:CentralizerSchema}
\vcenter{\hsize=0.4\textwidth\pspicture(0,0)(5.25,7)
\psset{linewidth=\pstlw,xunit=0.5,yunit=0.5,runit=0.5}
\psset{arrowsize=2pt 2,arrowinset=0.2}
%
%
\psline{-}(4,8.5)(6.5,8.5)
\psline{-}(6.5,8.5)(6.5,13.5)
\psline{-}(6.5,13.5)(4,13.5)
\psline{-}(4,13.5)(4,8.5)
%
%
\psbezier(5,12.5)(5,11.5)(4.5,12.5)(4.5,11.5)
\psbezier[border=1pt,bordercolor=white](4.5,12.5)(4.5,11.5)(5,12.5)(5,11.5)
\psbezier(5,11.5)(5,10.5)(4.5,11.5)(4.5,10.5)
\psbezier[border=1pt,bordercolor=white](4.5,11.5)(4.5,10.5)(5,11.5)(5,10.5)
\psbezier(5,10.5)(5,9.5)(4.5,10.5)(4.5,9.5)
\psbezier[border=1pt,bordercolor=white](4.5,10.5)(4.5,9.5)(5,10.5)(5,9.5)
\psbezier(5,12.5)(5,13)(5.5,13)(5.5,12.5)
\psbezier(4.5,12.5)(4.5,13.5)(6,13.5)(6,12.5)
\psbezier(5,9.5)(5,9)(5.5,9)(5.5,9.5)
\psbezier(4.5,9.5)(4.5,8.5)(6,8.5)(6,9.5)
\psline{-}(5.5,9.5)(5.5,12.5)
\psline{-}(6,9.5)(6,12.5)
\rput(7.5,9){\psframebox*[framesep=0pt]{${\mathfrak B}_n$}}
%
%
\psline{->}(3.5,8)(2.5,7)
\psline{->}(7.5,8)(8.5,7)
%
%
\psline{-}(0,1.5)(2.5,1.5)
\psline{-}(2.5,1.5)(2.5,6.5)
\psline{-}(2.5,6.5)(0,6.5)
\psline{-}(0,1.5)(0,6.5)
%
%
\psbezier(1,5.5)(1,4.5)(0.5,5.5)(0.5,4.5)
\psbezier(0.5,5.5)(0.5,4.5)(1,5.5)(1,4.5)
\psbezier(1,4.5)(1,3.5)(0.5,4.5)(0.5,3.5)
\psbezier(0.5,4.5)(0.5,3.5)(1,4.5)(1,3.5)
\psbezier(1,3.5)(1,2.5)(0.5,3.5)(0.5,2.5)
\psbezier(0.5,3.5)(0.5,2.5)(1,3.5)(1,2.5)
\psbezier(1,5.5)(1,6)(1.5,6)(1.5,5.5)
\psbezier(0.5,5.5)(0.5,6.5)(2,6.5)(2,5.5)
\psbezier(1,2.5)(1,2)(1.5,2)(1.5,2.5)
\psbezier(0.5,2.5)(0.5,1.5)(2,1.5)(2,2.5)
\psline{-}(1.5,2.5)(1.5,5.5)
\psline{-}(2,2.5)(2,5.5)
\rput(0,0.5){\psframebox*[framesep=0pt]{${\mathfrak B}_n\vert_{\grpGL}={\mathfrak S}_n$}}
\rput(3.25,4){\psframebox*[framesep=0pt]{$\not=$}}
%
%
%
\psline{-}(4,1.5)(7.5,1.5)
\psline{-}(7.5,1.5)(7.5,6.5)
\psline{-}(7.5,6.5)(4,6.5)
\psline{-}(4,1.5)(4,6.5)
%
%
\psline{-}(4.5,5)(4.5,5.5)
\psline{-}(5,5)(5,5.5)
\psarc(5,5){0.5}{180}{360}
\psarc(5.5,5){0.5}{180}{360}
\psarc(5.75,5){0.25}{0}{180}
\psbezier(5,4.5)(5,4.25)(5,4.25)(5,4)
\psbezier(5.5,4.5)(5.5,4.25)(4.5,4.25)(4.5,4)
\psarc(5,4){0.5}{180}{360}
\psarc(5.5,4){0.5}{180}{360}
\psarc(5.75,4){0.25}{0}{180}
\psbezier(5,3.5)(5,3.25)(5,3.25)(5,3)
\psbezier(5.5,3.5)(5.5,3.25)(4.5,3.25)(4.5,3)
\psarc(5,3){0.5}{180}{360}
\psarc(5.5,3){0.5}{180}{360}
\psarc(5.75,3){0.25}{0}{180}
\psbezier(5,5.5)(5,6)(6.5,6)(6.5,5.5)
\psbezier(4.5,5.5)(4.5,6.5)(7,6.5)(7,5.5)
\psbezier(5.5,2.5)(5.5,2)(7,1.5)(7,2.5)
\psbezier(5,2.5)(5,1.5)(6.5,2)(6.5,2.5)
\psline{-}(6.5,2.5)(6.5,5.5)
\psline{-}(7,2.5)(7,5.5)
\rput(6,0.5){\psframebox*[framesep=0pt]{$f_*({\mathfrak B}_n\vert_{\grpH_\pi})$}}
\rput(8.5,4.75){\psframebox*[framesep=0pt]{$f_*$}}
\psline{<-}(8,4)(9,4)
%
%
\psline{-}(9.5,1.5)(12,1.5)
\psline{-}(12,1.5)(12,6.5)
\psline{-}(12,6.5)(9.5,6.5)
\psline{-}(9.5,6.5)(9.5,1.5)
%
%
\psbezier(10.5,5.5)(10.5,4.5)(10,5.5)(10,4.5)
\psbezier[border=1pt,bordercolor=white](10,5.5)(10,4.5)(10.5,5.5)(10.5,4.5)
\psbezier(10.5,4.5)(10.5,3.5)(10,4.5)(10,3.5)
\psbezier[border=1pt,bordercolor=white](10,4.5)(10,3.5)(10.5,4.5)(10.5,3.5)
\psbezier(10.5,3.5)(10.5,2.5)(10,3.5)(10,2.5)
\psbezier[border=1pt,bordercolor=white](10,3.5)(10,2.5)(10.5,3.5)(10.5,2.5)
\psbezier(10.5,5.5)(10.5,6)(11,6)(11,5.5)
\psbezier(10,5.5)(10,6.5)(11.5,6.5)(11.5,5.5)
\psbezier(10.5,2.5)(10.5,2)(11,2)(11,2.5)
\psbezier(10,2.5)(10,1.5)(11.5,1.5)(11.5,2.5)
\psline{-}(11,2.5)(11,5.5)
\psline{-}(11.5,2.5)(11.5,5.5)
\rput(11,0.5){\psframebox*[framesep=0pt]{${\mathfrak B}_n\vert_{\grpH_\pi}$}}
\endpspicture
 
}
&
\vcenter{\hsize=0.4\textwidth\noindent Schematic mapping of a braid
element in $\frak{B}_n$, representing a knot or link, onto different representations.
The $\frak{S}_n$ representation is trivial, while we will show that the
$\sf{H}_\pi$ representation is not.}
\end{align}
The remainder of the paper is organised as follows. In  \S \ref{subsec:SymmX} we
introduce the character ring {\sf Char-GL} of finite dimensional polynomial
representations of the general linear group, in the guise of the ring of
symmetric functions $\Lambda(X)$ of symmetric polynomials in an alphabet $X$ of
countably many variables.  The Hopf algebra structure of $\Lambda(X)$ is
introduced and explained via the basic tangle diagram toolkit of \S
\ref{subsec:TangleTools}. In \S \ref{sec:CharHpi} we recall the results of
\cite{fauser:jarvis:king:wybourne:2005a} and explain the group character
branching rules needed to define {\sf Char-H$_\pi$}, and the modified
$\pi$-Newell Littlewood product rule (\S \S \ref{subsec:BranchingRules},
\ref{subsec:PiNewellLittlewood}). Associated with these products are 2-2
tangles, which are given diagrammatically in \S \ref{subsec:TanglesCrossings}
and shown to satisfy the braid relation. From the point of view of knots (\S
\ref{sec:KnotsnLinks}), these 2-2 tangles or crossings are the first part of the
knot alphabet, the braid relations themselves being equivalent to one of the
standard Reidemeister moves of knot isotopy. This connection is made
in \S \ref{subsec:KnotAlphabet}, and here the remaining elements of the knot
alphabet are defined. This leads to the main result of the paper:
\smallskip

\noindent\textbf{Theorem \ref{thm:MainTheorem}}: 
\textit{The character ring $\Lambda=\mbox{\sf Char-H}_{\pi}$ is a ribbon Hopf
algebra.}
\smallskip

For each partition $\pi$ the space $\Lambda\cong \CharHpi$ equipped with the
braid operators ${\texttt c}^\pi$, $({\texttt c}^\pi){}^{-1} =:
\overline{\texttt{c}}^\pi$ together with the morphisms   $\bcap$, $\dcup$,
$\bpcap$, $\dpcup$ and the canonical writhe element $Q_\pi$, is a ribbon Hopf
algebra.

With these ingredients at hand, in~\S~\ref{subsec:KnotInvariants} we 
identify `knot invariants' in this formalism. These are formed in 
the standard way by cutting and opening one or more strands of the 
closed braid representation of the knot or link, and so are formally 
ring endomorphisms, or operators on characters. Their diagrammatic 
evaluation is followed through systematically, with the result that 
with our present tool-kit, they turn out to be rather weak 
invariants. In the case of a simple knot, they merely report a 
measure of the writhe of the knot projection; for a link however, 
the invariants return the writhe of the individual components, as 
well as their mutual linking numbers. While this is but a small 
return for the large investment in mathematical structure leading to 
these final calculations, the underlying strategy is robust, and 
sheds original light on the background setting of algebraic and 
combinatorial aspects of representation theory ~\cite{barcelo:ram:1997a}. 

Finally we want to emphasise that the novelty of our approach lies not
in the rather trivial knot invariants presented here, but in the
fact that we use group-subgroup branchings, and that our method works in
the category of algebraic \grpGL-subgroups, which is more general than
that of semisimple Lie groups or algebras. There are several obvious
extensions which need to be pursued, with a possibility of gaining
better invariants. Among them are: $q$-deformations using Hall-Littlewood
or Macdonald functions and the like, generalization to noncommutative
symmetric functions, lifting our method from characters to
representation modules, analyzing a Drinfeld double of our Hopf
algebra, using graded modules.

\section*{Acknowledgement}
This work is the result of a collaboration over several years following
on from our paper \cite{fauser:jarvis:king:wybourne:2005a} with the late
Brian G Wybourne. BF gratefully acknowledges the Alexander von Humboldt
Stiftung for \emph{sur place} travel grants to visit the School of
Mathematics and Physics, University of Tasmania, and the University of
Tasmania for an honorary Research Associate appointment. Likewise PDJ
acknowledges longstanding support from the Alexander von Humboldt Foundation,
in particular for visits to the Max Planck Institute for Mathematics in
the Sciences, Leipzig. PDJ also acknowledges the Australian American
Fulbright Foundation for the award of a senior Fulbright scholarship.
RCK acknowledges support for part of this research from the Leverhulme
Foundation in the form of an Emeritus Fellowship.
Part of this work was presented at AUSTMS 2009, and BF thanks
Ruibin Zhang for support. We thank the DAAD for
financial support allowing a visit of RCK and PDJ to Erlangen, and we
are grateful for financial support from the Emmy-Noether Zentrum f\"ur
Algebra, at the University of Erlangen. The authors thank these
institutions for hospitality during our collaborative visits. Part of
this work was done under a `Research in Pairs' grant from the
Mathematisches Forschungsinstitut Oberwolfach, and the authors also wish
to record their appreciation of this award and the hospitality extended
to them in Oberwolfach. Finally, this work could not have been completed
without the generous support of the Quantum Computing Group, Department
of Computer Science, University of Oxford, for hosting a research visit.
\section{\CharGL and tangle diagrams}
\label{sec:CharGL}

\subsection{The ring of symmetric functions $\Lambda(X)$} 
We consider characters of finite-dimensional polynomial (tensor) representations
of the complex group $GL(n)$ of $n\times n$ nonsingular matrices, extended to a
ring over ${\mathbb Z}$ including formal subtraction as well as addition and
multiplication of characters. In the inductive limit $n\rightarrow \infty$ this
object \CharGL  is isomorphic to the ring of symmetric functions $\Lambda(X)$ on
an alphabet $X$ of countably many variables  $\{ x_1,x_2,x_3,\cdots \}$.
$\Lambda(X)$ has a canonical basis involving irreducible $GL(n)$ characters, the
Schur or $S$-functions $\{ s_\lambda \}_\lambda$ where $\lambda$ is an integer
partition $\lambda = (\lambda_1,\lambda_2, \cdots, \lambda_\ell)$ with
$\lambda_1 \ge \lambda_2 \ge \cdots \ge  \lambda_\ell >0$, with $\ell(\lambda) =
\ell$ the number of (positive) parts, and $|\lambda| = \lambda_1 + \lambda_2 +\cdots
\lambda_\ell$ the weight of the partition, written $\lambda \vdash |\lambda|$.
We follow \cite{macdonald:1979a} to which we refer for details; further
combinatorial and notational aspects will be introduced as required. Below we
turn to the attributes of $\Lambda(X)$ which play a   crucial role in our
development, namely its algebraic and Hopf-algebraic structures
\cite{fauser:jarvis:2003a}. Where no ambiguity arises, $\Lambda(X)$ will
occasionally be referred to simply as $\Lambda$ in what follows.
\label{subsec:SymmX} 

\subsection{Tangle diagram tool kit}\label{subsec:TangleTools} 
We wish to describe the algebraic properties of $\Lambda(X)$ in a pictorial way
using diagrams. A tangle diagram is a graph or decorated graph which represents
an algebraic statement. It contains zero or more pendant `input' edges, nodes
describing algebraic operations on whatever is being carried by the edges, and
finally pedant `output' edges (see for example
\cite{kuperberg:1991a,fauser:2002c}). In our convention algebraic operations
develop from top to bottom; directedness of the edges themselves has a different
significance, as we shall see. Here we illustrate the method while introducing
step-by-step the algebraic properties of $\Lambda(X)$.

In the simplest case of a single line or 1-1 tangle, the edge label (say $s_\lambda$) is unchanged, so we have the identity map:
\begin{align}
 \vcenter{\hsize=0.1\textwidth
\scalebox{0.6} 
{
\begin{pspicture}(0,-1.02)(0.02,1.02)
\psline[linewidth=0.04cm](0.0,1.0)(0.0,-1.0)
\end{pspicture} 
}

}
  &\Leftrightarrow
  &&
  {\sf Id}(s_\lambda)=s_\lambda 
  &&\textrm{(identity).}
\end{align}
By contrast the unit $\eta : \openZ\rightarrow \Lambda$ is an injection map from
the underling ring, say $\openZ$, into the ring of symmetric functions and is a
0-1 tangle:
\begin{align}
  \vcenter{\hsize=0.1\textwidth
\scalebox{0.6} 
{
\begin{pspicture}(0,-0.85)(0.325,0.85)
\psline[linewidth=0.05cm](0.15,0.575)(0.15,-0.825)
\psdots[dotsize=0.3,fillstyle=solid,dotstyle=o](0.15,0.675)
\end{pspicture} 
}

}
  &\Leftrightarrow
  &&
  \eta(1)=s_{0}
  && \textrm{(unit).}
\end{align}
These examples illustrate the convention that diagrams with unlabelled edges
specify maps, while edge labelling corresponds to specifying the action of maps
on elements.  0-$\!n$ tangles are injection maps and so have no input line, the
default action being on the scalar unit, $1$; conversely $n\!$-0 tangles are
scalar-valued, and so have no output line.

The symmetric function outer product is simply pointwise multiplication, and so
is a 2-1 tangle, ${\sf m}: \Lambda \otimes \Lambda \rightarrow \Lambda$:
\begin{align}
  \vcenter{\hsize=0.1\textwidth
\scalebox{0.6} 
{
\begin{pspicture}(0,-1.025)(1.225,1.025)
\psarc[linewidth=0.05](0.6,0.3){0.6}{-180.0}{0.0}
\psline[linewidth=0.05cm](0.0,0.3)(0.0,1.0)
\psline[linewidth=0.05cm](1.2,0.3)(1.2,1.0)
\psline[linewidth=0.05cm](0.6,-0.3)(0.6,-1.0)
\end{pspicture} 
}

}
  &\Leftrightarrow
  &&
  {\sf m}(s_\lambda\otimes s_\mu)=s_\lambda\cdot s_\mu
  =\sum_\nu c^\nu_{\lambda,\mu}s_\nu
  &&\textrm{(outer product).}
\end{align}
The coefficients $c^\nu_{\lambda,\mu}$ in the $S$-function basis are the famous
Littlewood-Richardson coefficients giving the multiplicity of irreducible parts
in the decomposition of a tensor product. The dual operation, a 1-2 tangle, is
the outer coproduct map $\Delta: \Lambda \rightarrow \Lambda \ot \Lambda$:
\begin{align}
  \vcenter{\hsize=0.1\textwidth
\scalebox{0.6} 
{
\begin{pspicture}(0,-1.025)(1.225,1.025)
\psarc[linewidth=0.05](0.6,-0.4){0.6}{-0.0}{180.0}
\psline[linewidth=0.05cm](0.6,0.2)(0.6,1.0)
\psline[linewidth=0.05cm](0.0,-0.4)(0.0,-1.0)
\psline[linewidth=0.05cm](1.2,-0.4)(1.2,-1.0)
\end{pspicture} 
}

}
  &\Leftrightarrow
  &&
  \Delta(s_\lambda)=\sum_{\mu,\nu} c^\lambda_{\mu,\nu}s_\mu \otimes s_\nu
  &&\textrm{(outer coproduct).}
\end{align}
which can be thought of as associating to a given symmetric function, a sum of
bilinears in symmetric functions from two distinct alphabets. A streamlined
notation for this sum is the Sweedler convention \cite{sweedler:1969a},
$\CO(s_\lambda) = \sum s_{\lambda_{(1)}} \ot s_{\lambda_{(2)}}$. These diagrams
show also that the juxtaposition of pendant edges is labelled by an element of
the tensor product of copies of $\Lambda$, in this case $\Lambda\ot \Lambda$, or
$\ot^n \Lambda$ for a tangle with $n$ input edges -- the most general tangle
diagram thus represents an algebraic statement in the tensor algebra
$T(\Lambda)$. 

Accompanying the outer coproduct is the   counit, a 1-0 tangle $\varepsilon :
\Lambda\rightarrow \openZ$, mapping a symmetric function to a ring element
(linear form):
\begin{align}
  \vcenter{\hsize=0.1\textwidth
\scalebox{0.6} 
{
\begin{pspicture}(0,-0.745)(0.32,0.75)
\psline[linewidth=0.05cm](0.15,0.725)(0.15,-0.475)
\psdots[dotsize=0.20040001,fillstyle=solid,dotstyle=o](0.15,-0.575)
\psdots[dotsize=0.3,fillstyle=solid,dotstyle=o](0.15,-0.575)
\end{pspicture} 
}

}
  &\Leftrightarrow
  &&
  \varepsilon(s_\lambda)=\delta_{\lambda,(0)}
  &&\textrm{(counit).}
\nonumber
\end{align}
An important adjunct to the above is another operation, that of $S$-function
\emph{skew}  which should be thought of as an endomorphism. This is the formal
dual of multiplication with respect to the Schur-Hall scalar product,
\begin{align}
  \vcenter{\hsize=0.17\textwidth\input{tangles/skew.tex}}
  &\Leftrightarrow
  &&
  s^\perp_\lambda(s_\mu)=\sum \la s_\lambda \vert s_\mu^{(1)}\ra s_\mu^{(2)}
  &&\textrm{(skew by $s_\lambda$).}
\nonumber
\end{align}
Defining alternatively $\langle {\sf D}(s_\lambda)s_\mu \vert s_\nu\rangle =$
$\langle s_\mu\vert s_\lambda s_\nu \rangle$, we have ${\sf D}(s_\lambda) = s_\lambda^\perp$ and the explicit form 
$s^\perp_\lambda(s_\mu) = \sum_\alpha c^\mu_{\lambda,\alpha} s_\alpha$.  
Because of the similarity to `division', the skew is often denoted 
$s_{\mu/\lambda} := s^\perp_\lambda(s_\mu)$.

A further structural element is the antipode map,
\begin{align}
  \vcenter{\hsize=0.1\textwidth
\scalebox{0.6} 
{
\begin{pspicture}(0,-1.025)(1.2509375,1.025)
\psline[linewidth=0.05cm](1.0809375,1.0)(1.0809375,-1.0)
\psdots[dotsize=0.3,fillstyle=solid,dotstyle=o](1.0809375,0.0)
\usefont{T1}{ptm}{m}{n}
\rput(0.54234374,0.305){$\antip$}
\end{pspicture} 
}

}
  &\Leftrightarrow
  &&
  \antip(s_\lambda)=(-1)^{\vert \lambda\vert}s_{\lambda'}
  &&\textrm{(antipode),}
\end{align}
related to the $\omega$-involution defined in \cite{macdonald:1979a}, which
serves to illustrate a final convention, that endomorphisms (linear operators)
are designated by in-line symbols, which still require edge labelling for their
action on elements to be specified. Here $\lambda'$ denotes the transposed
partition, formed by interchanging rows and columns of $\lambda$.

The setting in linear algebra implied by the elements described so far is
formalised by noting further that $\Lambda(X)$ is given the structure of a
Hilbert space, with orthonormal basis $\{s_\lambda \}_\lambda$, by the famous
Schur-Hall scalar product
\begin{align}
  \vcenter{\hsize=0.15\textwidth
\scalebox{0.6} 
{
\begin{pspicture}(0,-1.025)(2.025,1.025)
\psarc[linewidth=0.05](1.0,0.0){1.0}{-180.0}{0.0}
\psline[linewidth=0.05cm](0.0,0.0)(0.0,1.0)
\psline[linewidth=0.05cm](2.0,0.0)(2.0,1.0)
\end{pspicture} 
}

}
  &\Leftrightarrow
  &&
  \la s_\mu\mid s_\nu\ra=\delta_{\mu,\nu}
  &&\textrm{(Schur-Hall scalar product).}
\end{align}
This has a dual which injects the canonical element given by the sum over
paired basis vectors,
\begin{align}
  \vcenter{\hsize=0.17\textwidth
\scalebox{0.6} 
{
\begin{pspicture}(0,-1.025)(2.025,1.025)
\psarc[linewidth=0.05](1.0,0.0){1.0}{-0.0}{180.0}
\psline[linewidth=0.05cm](0.0,0.0)(0.0,-1.0)
\psline[linewidth=0.05cm](2.0,0.0)(2.0,-1.0)
\end{pspicture} 
}

}
  &\Leftrightarrow
  &&
  \sum_\lambda s_\lambda\otimes s_\lambda 
  &&\textrm{(Cauchy kernel $C[X,Y]$).}
\end{align}
Writing this in another way using two alphabets, we have the famous Cauchy
identity \cite{macdonald:1979a},
\begin{align}
  \sum_\lambda s_\lambda(X)  s_\lambda(Y) 
    &= \prod_{i,j} \frac{1}{(1-x_iy_j)} = C[X,Y],
\end{align}
whereas the dual Cauchy kernel is given by either of the following forms,
\begin{align*}
  \vcenter{\hsize=0.14\textwidth
\scalebox{0.6} 
{
\begin{pspicture}(0,-1.025)(3.181875,1.025)
\psarc[linewidth=0.05](1.0,0.0){1.0}{-0.0}{180.0}
\psline[linewidth=0.05cm](0.0,0.0)(0.0,-1.0)
\psline[linewidth=0.05cm](2.0,0.0)(2.0,-1.0)
\psdots[dotsize=0.3,fillstyle=solid,dotstyle=o](2.0,0.0)
\usefont{T1}{ptm}{m}{n}
\rput(2.5614061,0.405){$\antip$}
\end{pspicture} 
}

}
  &\Leftrightarrow
  &&
  {\sf Id}\ot {\sf S} \kern+0.15ex{\scriptstyle \circ} 
   \sum_\lambda s_\lambda\otimes s_\lambda 
   = \sum_\lambda (-1)^{\vert\lambda\vert} s_\lambda\otimes s_{\lambda'},
  &&\textrm{(Cauchy-Binet kernel),}
\nn
  \vcenter{\hsize=0.17\textwidth
\scalebox{0.6} 
{
\begin{pspicture}(0,-1.025)(3.1059375,1.025)
\psarc[linewidth=0.05](2.0809374,0.0){1.0}{-0.0}{180.0}
\psline[linewidth=0.05cm](1.0809375,0.0)(1.0809375,-1.0)
\psline[linewidth=0.05cm](3.0809374,0.0)(3.0809374,-1.0)
\psdots[dotsize=0.3,fillstyle=solid,dotstyle=o](1.0809375,0.0)
\usefont{T1}{ptm}{m}{n}
\rput(0.54234374,0.605){$\antip$}
\end{pspicture} 
}

}
  &\Leftrightarrow
  &&
  {\sf S}\ot{\sf Id} \kern +0.15ex{\scriptstyle \circ} 
   \sum_\lambda s_\lambda\otimes s_\lambda  
   = \sum_\lambda (-1)^{\vert\lambda\vert} s_{\lambda'} \otimes s_{\lambda},
  &&\textrm{(Cauchy-Binet kernel),}
\end{align*}\vskip-1.5ex
\begin{align} 
  \mbox{with} 
  &&
  \sum_\lambda (-1)^{\vert\lambda\vert} s_{\lambda'}(X)  s_\lambda(Y) 
  = \sum_\lambda (-1)^{\vert\lambda\vert} s_{\lambda}(X)  s_{\lambda'}(Y) = \prod_{i,j}{(1-x_iy_j)}, 
\end{align}
the latter being the Cauchy-Binet formula. A final example is the 2-2 tangle
representing the transposition or switch operator, which is simply
\begin{align}
  \vcenter{\hsize=0.14\textwidth
\scalebox{0.6} 
{
\begin{pspicture}(0,-1.025)(1.825,1.025)
\psline[linewidth=0.05cm](1.8,1.0)(0.0,-1.0)
\psline[linewidth=0.05cm](0.0,1.0)(1.8,-1.0)
\end{pspicture} 
}

}
  &\Leftrightarrow
  &&
  \textsf{sw}(s_\mu\otimes s_\nu)
  =s_\nu\otimes s_\mu
  &&\textrm{(switch).}
\end{align}
This tangle is planar and does not record any over or under information.

\subsection{\CharGL as a Hopf algebra}
\label{subsec:CharGLHopf}
We now give the properties satisfied by $\Lambda \cong \mbox{\CharGL}$
as a Hopf algebra, in the form of tangle diagrams. Firstly, the following
statements of the associativity of multiplication {\sf m} and
comultiplication $\CO$ are self-evident: 
\begin{align}
 \vcenter{\hsize=0.32\textwidth\input{tangles/associativity.tex}},
 &&
 \vcenter{\hsize=0.32\textwidth\input{tangles/coassociativity.tex}}.
\end{align}
It is useful to note that as a consequence, associativity and
co-associativity iterate to $n$-1 and 1-$n$ tangles which are
independent of the bracketing order, giving, effectively, diagrammatical
elements of star type. Moreover, the algebra and coalgebra are unital
and counital, respectively:
\begin{align}
 \vcenter{\hsize=0.23\textwidth\input{tangles/unitalProd.tex}},
&&
 \vcenter{\hsize=0.22\textwidth\input{tangles/counitalCoProd.tex}},
\end{align}
Also, in the particular case of $\Lambda$, multiplication {\sf m} is
commutative and and comultiplication $\CO$ is cocommutative:
\begin{align}
 \vcenter{\hsize=0.18\textwidth\input{tangles/commutativityProd.tex}},
&&
 \vcenter{\hsize=0.18\textwidth\input{tangles/cocommutativityCoProd.tex}},
\end{align}
The bialgebra property asserts
\begin{align}
 \vcenter{\hsize=0.3\textwidth\input{tangles/axiomBialgebra.tex}}
 &\Leftrightarrow
 &&
 \Delta(fg) = \Delta(f)\cdot\Delta(g),
\end{align}
where $f$ and $g$ are any symmetric functions. The operator representing the
tangle giving the right-hand side is ${\sf m}\ot{\sf m} \comp 1 \ot {\sf sw} 
\ot 1\comp \Delta \ot \Delta (f\ot g)$. If we write generically
\cite{sweedler:1969a} $\Delta f = \sum f_{(1)} \otimes f_{(2)}$, then the
axiom reads
\begin{align}
  \Delta(f \cdot g) 
    &= \sum f_{(1)} \cdot g_{(1)} \otimes \sum f_{(2)} \cdot g_{(2)}.
\end{align}
Finally, augmenting the bialgebra property, the antipode map satisfies the
compatibility conditions
\begin{align}
 \vcenter{\hsize=0.3\textwidth\input{tangles/axiomAntipode.tex}}
\end{align}
which state that the (idempotent) antipode is the convolutive inverse of the
identity map. Finally we have
\mybenv{Theorem} \textbf{The outer Hopf algebra of symmetric functions:}
\label{thm:OuterHopf}
The sextuple $(\Lambda, {\sf m}, \CO, \antip,\eta,\varepsilon)$ with the
above bialgebra, (co)unit and antipode axioms, is a commutative,
co-commutative Hopf algebra, the outer Hopf algebra of symmetric functions.
\myeenv
\noindent
\textbf{Proof}:
see for example \cite{thibon:1991a,thibon:1991b,fauser:jarvis:2003a}.
\qed

To end this section, the following display exhibits several tangle
relations expressing algebraic identities that are easily checked
in the \CharGL case, where the under and over crossings are identical.

\begin{align}\label{fig:GLrelations}
 \vcenter{\hsize=0.12\textwidth
\scalebox{0.6} 
{
\begin{pspicture}(0,-1.025)(2.425,1.025)
\psarc[linewidth=0.05](0.6,0.0){0.6}{-0.0}{180.0}
\psline[linewidth=0.05cm](0.0,0.0)(0.0,-1.0)
\psarc[linewidth=0.05](1.8,0.0){0.6}{-180.0}{0.0}
\psline[linewidth=0.05cm](2.4,0.0)(2.4,1.0)
\end{pspicture} 
}

}
 \simeq
 \vcenter{\hsize=0.05\textwidth
\scalebox{0.6} 
{
\begin{pspicture}(0,-1.02)(0.02,1.02)
\psline[linewidth=0.04cm](0.0,1.0)(0.0,-1.0)
\end{pspicture} 
}

}
 \simeq
 \vcenter{\hsize=0.12\textwidth
\scalebox{0.6} 
{
\begin{pspicture}(0,-1.025)(2.425,1.025)
\psarc[linewidth=0.05](1.8,0.0){0.6}{-0.0}{180.0}
\psline[linewidth=0.05cm](0.0,1.0)(0.0,0.0)
\psarc[linewidth=0.05](0.6,0.0){0.6}{-180.0}{0.0}
\psline[linewidth=0.05cm](2.4,-1.0)(2.4,0.0)
\end{pspicture} 
}

}
 &&
 \vcenter{\hsize=0.45\textwidth\noindent \textbf{R0:} The zeroth or topological 
          Reidemeister move, also the closure of the tangle category.}
 \\ \label{reidemeisterI}
 \vcenter{\hsize=0.12\textwidth
\scalebox{0.6} 
{
\begin{pspicture}(0,-1.025)(1.925,1.025)
\psarc[linewidth=0.05](1.3,0.0){0.6}{-0.0}{135.0}
\psline[linewidth=0.05cm](0.0,-1.0)(0.88,0.44)
\psline[linewidth=0.05cm](0.0,1.0)(0.82,-0.34)
\psarc[linewidth=0.05](1.3,0.0){0.6}{213.69006}{0.0}
\end{pspicture} 
}

}
 \simeq
 \vcenter{\hsize=0.05\textwidth
\scalebox{0.6} 
{
\begin{pspicture}(0,-1.02)(0.02,1.02)
\psline[linewidth=0.04cm](0.0,1.0)(0.0,-1.0)
\end{pspicture} 
}

}
 \simeq
 \vcenter{\hsize=0.12\textwidth\input{tangles/reidemeister1r.tex}}
 &&
 \vcenter{\hsize=0.45\textwidth\noindent \textbf{R1:} The first Reidemeister 
          move.}
 \\
 \vcenter{\hsize=0.1\textwidth\input{tangles/reidemeister2l.tex}}
 \simeq
 \vcenter{\hsize=0.09\textwidth
\scalebox{0.6} 
{
\begin{pspicture}(0,-1.02)(1.12,1.02)
\psline[linewidth=0.04cm](0.0,-1.0)(0.0,1.0)
\psline[linewidth=0.04cm](1.1,-1.0)(1.1,1.0)
\end{pspicture} 
}

}
 \simeq
 \vcenter{\hsize=0.1\textwidth\input{tangles/reidemeister2r.tex}}
 &&
 \vcenter{\hsize=0.45\textwidth\noindent \textbf{R2:} The second Reidemeister 
          move.}
 \\
 \label{fig:GLrelations2}
 \vcenter{\hsize=0.14\textwidth\input{tangles/reidemeister3l.tex}}
 \simeq
 \vcenter{\hsize=0.16\textwidth\input{tangles/reidemeister3r.tex}}
 &&
 \vcenter{\hsize=0.45\textwidth\noindent \textbf{R3:} The third Reidemeister 
          move, also quantum Yang-Baxter equation.}
\end{align}
The first two relations are `straightening rules' which allow loops to be
removed; the second two statements mean that the switch {\sf sw} has all
the properties of an elementary transposition operator in the symmetric
group, if the permutations (on $n$ objects say) are realised as
$n\!$-$\!n$ tangles representing corresponding line shuffles. Thus the
third diagram asserts that elementary transpositions are involutive,
while the fourth diagram gives the standard exchange relation for
neighbouring transpositions in the presentation of the symmetric group:
\begin{align}\label{fig:Snrelations}
{\mathfrak S}_n = \left.\langle\right. s_i, i \in \{1,2,\cdots,n\}: s_i^2 = {\sf Id}, s_i s_j = s_js_i, |i-j|\ge 2,
s_is_{i+1}s_i = s_{i+1}s_is_{i+1} \left.\rangle\right.
\end{align}
under the morphism $\mathfrak{B}_n\rightarrow \mathfrak{S}_n$ under
which each braid generator $b_i$ maps to the corresponding
transposition $s_i$.

These relations serve as a template for the idea that tangle diagrams
are subject to rearrangement by simplification rules. This theme will
be developed with the subgroup Hopf algebras to be introduced in
later sections, and will of course be replaced by statements enabling
manipulations with nontrivial braid operators $\mathsf{c}_{i,j}$, when
the correct identifications are given.   

\section{The character rings {\sf Char-H$_\pi$}}
\label{sec:CharHpi}
For certain matrix subgroups of $GL(n)$, the characters can be handled by an
extension of the symmetric function methods in $\Lambda(X)$ used for \CharGL,
and these form the subject of the present section (see \S
\ref{subsec:BranchingRules} below). We frequently illustrate the results by the
special case of the \emph{classical} matrix subgroups, the  orthogonal and
symplectic groups $O(n)$ and $Sp(n)$, respectively (with $n$ even in the latter
case). The general class of subgroups $H_\pi(n)$ which we consider includes not
only, for example, odd-dimensional symplectic groups \cite{proctor:1988a}, and
orthogonal groups with singular metric, but generically non-semisimple matrix
groups and even discrete groups \cite{fauser:jarvis:king:wybourne:2005a}.
However, the tangle notation for $H_\pi(n)$ is more involved (see \S
\ref{subsec:TanglesCrossings}).   

\subsection{Group branching rules}
\label{subsec:BranchingRules}
The character rings of the orthogonal and symplectic groups can be treated with
symmetric function methods \cite{fauser:jarvis:king:2007c} provided care is
given to the way in which the conjugacy classes (parametrised by eigenvalues of
the group matrices) are handled. In the inductive limit, \CharO and \CharSp are,
as linear spaces, isomorphic copies of \CharGL, whereas the product rule for
their characters is different. The isomorphism of spaces is called the
group-subgroup branching rule, and reflects at the character level the
decomposition of irreducible representations, on restriction to a subgroup. The
multiplication of characters (corresponding to the reduction of a tensor
product) is called the Newell-Littlewood rule
\cite{newell:1951a,littlewood:1940a}. Similarly for the groups $H_\pi(n)$ in the
inductive limit, we shall introduce the character rings \CharHpi.  In the next
subsections we give the $\pi$-group-subgroup branching rules, and in the
following subsection we give the $\pi$-Newell-Littlewood product rule. 

\subsubsection{$S$-function series and plethysms}
In order to introduce the group-subgroup branching rule, we first turn to
systematics of $S$-function series \cite{macdonald:1979a}. These are formal
infinite sums of symmetric functions of a specific type (often involving
partitions of a particular shape if referred to the basis of $S$-functions)
whose elements are used term-wise in multiplication and skewing operations ( \S
\ref{subsec:TangleTools}). More correctly, the ring $\Lambda$ is extended to
$\Lambda[[t]]$, with the convention that if the symmetric functions appearing in
the coefficient ${[}t^n{]}$ are graded by partition weight, then the
indeterminate $t$ is redundant and is usually omitted.

The ring $\Lambda$ has various linear and multiplicative bases whose elements
are succinctly given by generating series. The most useful are 
\begin{align}
M_t = & \, \prod_i \frac{1}{1-tx_i} = \sum_n h_n t^n, \nonumber \\
L_t = & \, \prod_i {(1-tx_i)} = \sum_n e_n (-t)^n, \nonumber \\
P_t = & \, -\ln M_t := \sum_n p_n t^n/n, \nonumber \\
\mbox{where} \qquad 
p_n(X) = & \, \sum_i x_i^n, \nonumber \\
h_n(X) = & \, \sum_{1_1 \le i_2 \le \cdots \le i_n} x_{i_1} x_{i_2} \cdots x_{i_n},
\nonumber \\
e_n(X) = & \, \sum_{1_1 < i_2 < \cdots < i_n} x_{i_1} x_{i_2} \cdots x_{i_n}
\nonumber
\end{align}
are the power sum, complete and elementary symmetric functions, respectively.
The complete symmetric functions are in fact $S$-functions for one part
partitions, $h_n \equiv s_{(n)}$, and for partitions with each part at
most 1 (Young diagrams with one column, the transpose of the one part
partitions), $e_n \equiv s_{(1^n)}$, respectively. In general each
Schur function can be written
\begin{align}
  s_\lambda(X) 
    &= \sum_T x^T,
\end{align}
where $T$ is a monomial in $x_i$ derived from the entries in semistandard
tableaux $F_\lambda$ filling the Young frame or Ferrers diagram of $\lambda$.
For example, $s_{(2,1)}(X) =  \sum_{i\ne j} x_i^2 x_j  + 
2 \sum_{i<j<k} x_i x_j x_k $. In contexts where the alphabet is unambiguous, the
Littlewood convention $s_\lambda(X) = \{\lambda\}(X)$ or simply $s_\lambda
= \{\lambda \}$ is sometimes useful for Schur functions, in order to emphasize
the specific partition and if the alphabet is understood. Thus for example
$s_{(2,1)}(X)$ can be written as $\{2,1\}$.

In the following we shall also be interested in cases of the operation of
symmetric function \emph{plethysm} which is defined as
follows~\cite{fauser:jarvis:king:2007c}. For $\lambda$ a symmetric function
let $Y = \{x^T\}_{T\in F_\lambda}$ be the alphabet corresponding to the
monomials $x^T$. Then for any symmetric function $f$, the plethysm of $f$ by
$\lambda$ is  $f[s_\lambda](X) := f(Y)$. In particular we can evaluate
$s_\mu[s_\lambda](X)$ in this way. Applied to the above series, for a fixed
partition $\pi$, consider the semistandard tableaux $T\in F_\pi$  as above.
We define
\begin{align}
M_\pi \equiv M[s_\pi] = & \, \prod_{T \in F_\pi} \frac{1}{(1-x^T)},
\qquad   L_\pi \equiv L[s_\pi] = \prod_{T \in F_\pi} {(1-x^T)};    \nonumber \\
\mbox{thus}\qquad \qquad \qquad M_{\{2\}} = & \, \prod_{i \le j } \frac{1}{(1-x_i x_j)},
\qquad   L_{\{2\}} = \prod_{i \le j } {(1-x_i x_j)}, \nonumber \\
\qquad M_{\{1,1\}} = & \, \prod_{i < j } \frac{1}{(1-x_i x_j)},
\qquad   L_{\{1,1\}} = \prod_{i < j } {(1-x_i x_j)}. \nonumber 
\end{align}
Standard notation for the above is \cite{littlewood:1940a}
$M_{\{2\}}=D = \sum_\delta s_\delta$, $L_{\{2\}} = C = 
\sum_\gamma (-1)^{\frac 12\vert \gamma\vert} s_\gamma$, 
$M_{\{1,1\}}=B = \sum_\beta s_\beta$,
$L_{\{1,1\}} = A = \sum_\alpha (-1)^{\frac 12\vert \alpha\vert} s_\alpha$.
Here $\{\delta\}$ is the set of partitions with each part (row) even;
$\{\gamma\}$ is the set of partitions whose principal hooks (nested, inverted
L-shaped strips down the diagonal) have the shape $(a_i+1, 1^{b_i})$ with
$a_i = b_i+1$ (arm length exceeds leg length by 1); $\beta$ is the set of
partitions $\delta'$, and $\alpha$ is the set of $\gamma'$. 

For higher rank plethysms we have for example
\begin{align}
  M_{\{3\}} =  \prod_{i \le j \le k} \frac{1}{(1-x_i x_jx_k)}, \quad  & \, 
  L_{\{3\}} =  \prod_{i \le j \le k} {(1-x_i x_jx_k)}, \nonumber\\
   M_{\{2,1\}} = 
\prod_{i \ne j } \frac{1}{(1-x_i^2 x_j)} \prod_{i < j < k}\frac{1}{(1-x_i x_jx_k)}, 
 \quad  & \, L_{\{2,1\}} =   \prod_{i \ne j } {(1-x_i^2 x_j)} \prod_{i < j < k}{(1-x_i x_jx_k)}. 
\end{align}                                                                                                                                                               
For $\pi$ of rank 3 and higher, the first few terms in the series $M_{\{3\}}$,
$M_{\{2,1\}}$, $\cdots$  can be evaluated explicitly, but there is no known
systematic description of the patterns occurring at arbitrary degree.

\subsubsection{Group branching rules for $\pi$-characters.}
As mentioned above, at the character ring level, the group-subgroup branching
rule is a linear space isomorphism. In the case of the orthogonal and symplectic
subgroups of $GL(n)$, in the $S$-function basis, the images of $s_\lambda$ are
denoted $o_\lambda$ and $sp_\lambda$, respectively (Schur functions of
orthogonal and symplectic type) and are irreducible characters of the respective
subgroups. The branching isomorphisms are denoted $/D$ and $/B$, respectively:
\mybenv{Theorem} \textbf{Group branching rules -- orthogonal and symplectic
groups.}
\begin{align}
/D: \mbox{\CharGL}  \rightarrow \mbox{\CharO}, \qquad & \, s_\lambda \mapsto o_{\lambda/D};
\nonumber \\
/D^{-1} \equiv /C: \mbox{\CharO}  \rightarrow \mbox{\CharGL}, \qquad & \, o_\lambda \mapsto s_{\lambda/C};
\nonumber \\
/B: \mbox{\CharGL} \rightarrow \mbox{\CharSp}, \qquad & \, s_\lambda \mapsto sp_{\lambda/B};
\nonumber \\
/D^{-1} \equiv /C: \mbox{\CharSp} \rightarrow \mbox{\CharGL}, \qquad & \, sp_\lambda \mapsto s_{\lambda/A}.
\nonumber 
\end{align}
\myeenv
\noindent\textbf{Proof}: See for example Littlewood~\cite{littlewood:1940a}.
The fact that the inverse maps are also series branchings by the inverse
series is due to the distributivity law for the symmetric function skew
product, namely $f/(gh) = (f/g)/h$, which follows trivially from the duality
between skew product and outer multiplication. See
also~\cite{fauser:jarvis:king:2007c}.
\qed

Tangle diagrams for the group branching laws (skewing by symmetric function
series) are introduced as in \S~\ref{subsec:TangleTools} above. For example,
the following manipulations establish the inverse branchings, for any series
$W$ and $Z$ such that $WZ=1$, implementing $\big( s_\lambda  /W\big)/Z = 
s_\lambda  /(WZ) \equiv  s_\lambda$:
\begin{align}
  \vcenter{\hsize=0.15\textwidth\input{tangles/skewBranching1.tex}}
  \simeq
  \vcenter{\hsize=0.15\textwidth\input{tangles/skewBranching2.tex}}
  \simeq
  \vcenter{\hsize=0.15\textwidth\input{tangles/skewBranching3.tex}}
  \simeq
  \vcenter{\hsize=0.15\textwidth
\scalebox{0.6} 
{
\begin{pspicture}(0,-2.525)(1.375,2.525)
\psline[linewidth=0.05cm](0.15,1.1)(0.15,-0.1)
\psdots[dotsize=0.20040001,fillstyle=solid,dotstyle=o](0.15,-0.2)
\psdots[dotsize=0.3,fillstyle=solid,dotstyle=o](0.15,-0.2)
\psarc[linewidth=0.05](0.75,1.1){0.6}{-0.0}{180.0}
\psline[linewidth=0.05cm](0.75,1.7)(0.75,2.5)
\psline[linewidth=0.05cm](1.35,1.1)(1.35,-2.5)
\end{pspicture} 
}

}
\end{align}

Finally we turn to the general $H_\pi$ groups and their character rings
\CharHpi. Fix a partition $\pi$ and a complex tensor (an element of $\ot^{\vert
\pi \vert}({\mathbb C^n})$) of Young symmetry type $\pi$. Define $H_\pi(n)$ to
be the subgroup of the group $GL(n)$ of nonsingular complex $n\times n$ matrices
which leave invariant the fixed tensor under the natural action of $GL(n)$
induced by that on the fundamental representation ${\mathbb C}^n$. Clearly this
group depends critically on the canonical form of the fixed tensor, and may
indeed be trivial, or possibly discrete. In general, however, it will be a
certain algebraic subgroup of $GL(n)$ and -- except in the case of the classical
orthogonal and symplectic subgroups -- will be non-semisimple (as mentioned
above, including for example, symplectic groups in odd dimensions, orthogonal
groups with singular metrics). The ring \CharHpi (over ${\mathbb Z}$) is an
isomorphic copy of $\Lambda$ consisting of formal characters of finite
dimensional, in general indecomposable, representations of $H_\pi(n)$ in the
inverse limit, defined via branching maps
\cite{fauser:jarvis:king:wybourne:2005a}. In analogy with the orthogonal and
symplectic cases, we introduce symmetric functions of type $H_\pi$ denoted
$s^{(\pi)}_\lambda$, defined via branching maps as follows:
\mybenv{Definition}\textbf{Symmetric functions of type $H_\pi$}
\begin{align}
/M_\pi: \mbox{\CharGL}  \rightarrow \mbox{\CharHpi}, \qquad & \, s_\lambda \mapsto s^{(\pi)}_{\lambda/M_\pi};
\nonumber \\
/L_\pi : \mbox{\CharHpi}  \rightarrow \mbox{\CharGL}, \qquad & \, s^{(\pi)}_\lambda \mapsto s_{\lambda/L_\pi}.
\nonumber 
\end{align}
\myeenv

\subsection{$\pi$-Newell-Littlewood product and associated tangles}
\label{subsec:PiNewellLittlewood}
We now turn to the product rule for subgroup characters in the rings \CharHpi, which generalises that for the orthogonal and symplectic groups, \CharO and \CharSp. For the latter cases, the product of characters corresponds to the decomposition of a tensor product of irreducible representations, and in terms of $S$-functions is called the Newell-Littlewood rule,
\mybenv{Corollary}\textbf{Newell-Littlewood rule}:\\
\begin{align}
o(\lambda)  o(\mu)= & \,  \sum_\alpha o(\lambda/\alpha \cdot \mu/\alpha), \qquad
sp(\lambda) sp(\mu)= \sum_\alpha sp(\lambda/\alpha \cdot \mu/\alpha), \nonumber
\end{align}
\textbf{Proof}: See Newell \cite{newell:1951a}, and also Littlewood \cite{littlewood:1940a}. This is a special case of the $\pi$-Newell-Littlewood rule below (Theorem \ref{thm:piNewellLittlewood}) proved in \cite{fauser:jarvis:king:wybourne:2005a}.
\mbox{}\\
\mbox{}\hfill $\Box$

The generalisation to the product of characters in \CharHpi requires a
discussion of the Hopf structure of $s_\pi$. For clarity of writing, we
write $s(\lambda)$ or $\{ \lambda \}$, and $o(\mu)$, $sp(\nu)$,
$s^{(\pi)}(\alpha)$, to denote Schur functions, and symmetric functions
of orthogonal, symplectic and $H_\pi$ type, in $\Lambda$, respectively.
Where no confusion arises, enclosing braces ${\{} \cdot {\}}$ on Schur
functions can be omitted (where it is clear that expressions are evaluated
in the $S$-function basis, as opposed to acting on general elements
$f,g,\cdots$)\footnote{%
   Concrete instances such as ${\{}2,1{\}}$ still require braces to
   distinguish $S$-functions from the corresponding partitions, in
   this case $(2,1)$.}.
In the same vein, Schur function plethysms can be abbreviated, for
example $\alpha{[}\beta{]}$. Let $\{ {\{}\pi'_{(1)}{\}}\ot
{\{}\pi'_{(2)} {\}}\}$ be the list \emph{with repetition} of entries
of the cut outer coproduct $\Delta^\prime(\{\pi\})$ of ${\{}\pi{\}}$, that is,
\begin{align}
\Delta\big( \{ \pi \}\big) -  \,  \{ \pi \}  \ot 1 - 1 \ot \{ \pi \}
  &=  \Delta^\prime(\{\pi\}) = \sum \{ \pi'_{(1)} \} \ot \{ \pi'_{(2)} \}, \nonumber \\
\mbox{or simply} \qquad 
\Delta(\pi) - \, \pi  \ot 1 - 1 \ot \pi 
  &=  \Delta^\prime(\pi) = \sum  \pi'_{(1)}  \ot  \pi'_{(2)} . \nonumber 
\end{align}
Let the cardinality of this list be $p=\vert \Delta^\prime \pi \vert$.
Let $\{\alpha_k\}_{k=1}^p$ be a set of $p$ partitions, and denote by
$\sum_\alpha = \sum_{\alpha_1,\alpha_2,\cdots, \alpha_p}$ the
summation over all such $p$-tuples of partitions. We have

\mybenv{Theorem}\textbf{Generalised Newell-Littlewood rule}:\\
\label{thm:piNewellLittlewood}
\begin{align}
{\sf m}_\pi \big(s^{(\pi)}(\lambda) , s^{(\pi)}(\mu)\big) = s^{(\pi)}(\lambda) \piprod s^{(\pi)}(\mu)= & \,  \sum_{\alpha_k} 
s^{(\pi)}\big( \lambda/\prod_{k=1}^p \alpha_k[\pi'_{(1)}] \cdot 
\mu/\prod_{k=1}^p \alpha_k[\pi'_{(2)}] \big), \nonumber
\end{align}
\textbf{Proof}: See \cite{fauser:jarvis:king:wybourne:2005a}. \mbox{}\\
\mbox{}\hfill $\Box$

The Newell-Littlewood rule and its $\pi$-generalisation can be explicitly constructed via the above decompositions, at least for concrete cases. Firstly, note that the orthogonal and symplectic groups are, by definition, matrix subgroups of $GL(n)$ which leave invariant a bilinear form which is symmetric, or antisymmetric, respectively, so that $\pi = \{ 2 \}$ or $\pi = \{ 1,1\}$.
Noting that 
\[
\CO(\{2\}) = \{2\} \ot 1 + 1 \ot  \{2\}  + \{1\} \ot \{1\} ,
\qquad
\CO(\{1,1\}) = \{1,1\} \ot 1 + 1 \ot  \{1,1\}  + \{1\} \ot \{1\} ,
\]
we have in both cases $\Delta'(\{\pi\}) = \{1\} \ot \{1\} $, so there
is only one summand $\alpha$, and moreover the plethysm is trivial,
$\{\alpha{[}1{]}\} \equiv \{\alpha \}$ so the standard Newell-Littlewood
result is recovered. 

For $\vert \pi\ \vert > 2$ however there is more than one part in
$\Delta'(\{\pi\})$ and so summands $\alpha_1$, $\alpha_2$, $\cdots$ $\alpha_p$ are required, and at least some of these will come with nontrivial plethysms if parts of $\Delta(\{ \pi\})$ are of rank 2 or larger. For example.
\[
\CO(\{3\}) = \{3\} \ot 1 + 1 \ot  \{3\} +\{1\} \ot  \{2\} + \{2\} \ot \{1\} ,
\]
so that with 2 summands $ \alpha_1, \alpha_2$ or $\alpha$, $\beta$ say,
\[
s^{(3)}(\lambda) \piprod s^{(3)}(\mu)
= \sum_{\alpha,\beta} s^{(3)}\big( \lambda/(\alpha\!\!\cdot\! \!\beta[2]) \cdot \mu/(\alpha[2]\!\!\cdot\!\! \beta) \big).
\]

Using tangle diagrams, these deformed products are readily illustrated. In the
orthogonal and symplectic cases we have 
\begin{align}
  \vcenter{\hsize=0.4\textwidth\input{tangles/slicedNewellLittlewood.tex}}
  &\Leftrightarrow
  \begin{array}{l}
     \lambda \ot\mu \\
     \lambda_{(1)}\ot \lambda_{(2)}\ot\mu_{(1)}\ot\mu_{(2)} \\
     \sum_\alpha\lambda_{(1)}\ot \lambda_{(2)}\ot\mu_{(1)}\ot\mu_{(2)} \ot \alpha \ot \alpha \\ 
     \sum_\alpha\lambda_{(1)}\ot \mu_{(1)}\ot \lambda_{(2)} \ot \mu_{(2)}\ot \alpha \ot \alpha \\ 
     \sum_\alpha\lambda_{(1)}\ot \mu_{(1)}\ot \lambda_{(2)} \ot \la \mu_{(2)}\vert \alpha\ra \alpha\\ 
     \sum_\alpha\lambda_{(1)}\cdot \mu_{(1)}\ot \lambda_{(2)} \ot \la \mu_{(2)}\vert \alpha\ra \alpha\\ 
     \sum_\alpha\lambda_{(1)}\cdot \mu_{(1)}\la \lambda_{(2)} \vert \alpha \ra \la \mu_{(2)}\vert \alpha\ra 
  \end{array}.
\end{align}
Summing over $\alpha$ and using orthonormality of the $S$-functions, the last line becomes 
$\sum \lambda_{(1)}\cdot \mu_{(1)}\la \lambda_{(2)}\vert \mu_{(2)}\ra$ or $ \sum_\alpha \lambda/\alpha \cdot \mu/\alpha$, thus 
yielding the required form of the Newell-Littlewood rule, here denoted $\lambda \piprod \mu$. In the
general case, the Cauchy kernel is replaced by a more complicated 0-2 tangle
which injects the required summands with appropriate plethysms. Calling the
Cauchy kernels $r_{\{2\}}=r_{\{1,1\}}$, we have for $\pi = \{3\}$ the
corresponding operator $r_{\{3\}}$, where
\begin{align}
  \vcenter{\hsize=0.17\textwidth
\scalebox{0.6} 
{
\begin{pspicture}(0,-1.025)(2.025,1.025)
\psarc[linewidth=0.05](1.0,0.0){1.0}{-0.0}{180.0}
\psline[linewidth=0.05cm](0.0,0.0)(0.0,-1.0)
\psline[linewidth=0.05cm](2.0,0.0)(2.0,-1.0)
\end{pspicture} 
}

}
  &\Leftrightarrow
  && 
  r_{\{2\}}=r_{\{1,1\}}~::~  1 \mapsto 
   \left. \sum\right._\alpha \alpha \ot \alpha,
  \\
  \vcenter{\hsize=0.2\textwidth\input{tangles/rPi3.tex}}
  &\Leftrightarrow
  && 
  r_{\{3\}}~::~ 1 \mapsto \, \sum_{\alpha, \beta} 
     \alpha[2] \cdot \beta \ot \alpha \cdot \beta[2]. 
\end{align}
In general the 0-2 tangle $r_\pi$ is a convolutive product of
$p=\vert\Delta^\prime \pi\vert $ Cauchy kernels, whose downward lines are modified by the insertion of
plethysms coming from the corresponding cut coproduct parts of $\pi$.

\subsection{Crossings and the braid relation}
\label{subsec:TanglesCrossings}
Having identified a large class of nontrivial new product rules on $\Lambda$
arising from the decomposition of characters of type ${\sf H}_\pi$ (with each
partition $\pi$ of rank $>2$ giving a distinct multiplication) we explore in
this subsection the further ramifications of this structure. As explained above,
the central feature of the deformed multiplication is that it is an appropriate
convolution of ordinary multiplication, with a type of 0-2 tangle, $r_\pi$
(which is itself a convolution of modified Cauchy kernels).  Another object
closely related to $r_\pi$ is a 2-2 tangle on $\Lambda \ot \Lambda$, denoted
$R_\pi$. It is the task of this subsection to exploit the properties of $r_\pi$
in order to investigate $R_\pi$. We follow \cite{kassel:1995a,turaev:1994a} (see
also \cite{ohtsuki:2002a}). The outcome will be that $R_\pi$ can be regarded as
a new type of crossing, and indeed satisfies the quantum Yang-Baxter relation,
or when composed with the plain switch {\sf sw}, the braid relation. These
findings will provide the starting point for exhibiting the full alphabet of
knot moves~(\S~\ref{sec:KnotsnLinks} below). 

\mybenv{Definiton}\label{def:rpi}
\textbf{Generalised Cauchy kernel $r_\pi$ and Cauchy scalar $Q_\pi$}:
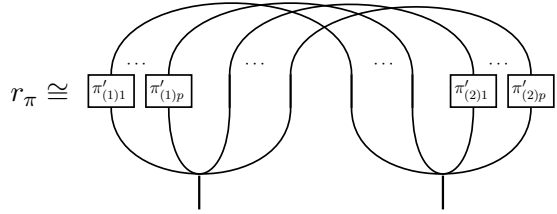
\begin{align}
  &&&&r_\pi  &\cong\hskip-4ex
    \vcenter{\hsize=0.45\textwidth\input{tangles/rPiGeneral.tex}}
  \nonumber\\
  \mbox{where}
  &&&&
  r_{\pi} &= \sum
\alpha^\pi_{(1)} \ot \alpha^\pi_{(2)}:=\sum_\alpha \prod_{k=1}^p \alpha_k[\pi'_{(1)}] 
               \ot               \prod_{k=1}^p \alpha_k[\pi'_{(2)}],
\label{eqn-willow}
\end{align}
where the boxed operators $\pi^\prime_{(n)}$ on downward lines signify that the appropriate ${[}\pi^\prime_{(n)}{]}$ plethysms 
are to be applied to the running variables $\alpha_1, \alpha_2,\cdots,\alpha_p$.
\myeenv
Here and in the sequel we adopt the convenient short-hand notation
$r_\pi =  \sum_\alpha \alpha^\pi_{(1)} \ot \alpha^\pi_{(2)}$ for the
cut co-product-derived Sweedler sums occurring in the coscalar product;
in Theorem \ref{thm:piNewellLittlewood} this convention would instead
lead to $\sum_\alpha \cdot/\alpha^\pi_{(1)} \ot \cdot / \alpha^\pi_{(2)}$.
In this notation the following scalar element or `tadpole', regarded as
a 0-1 tangle\footnote{Note that the 0-2 and 0-1 tangles $r_\pi$ and
$Q_\pi$ are technically homomorphisms, so could also have been defined
as $r_\pi(1)$ and $Q_\pi(1)$, respectively.},
arises as the multiplicative closure of the bottom lines of $r_\pi$,
\begin{align}
\label{eq:QpiDefinition}
Q_\pi = & \,  \sum_\alpha \alpha^\pi_{(1)} \cdot \alpha^\pi_{(2)}.
\end{align}

The fact that the $\pi$-modified Newell-Littlewood rule gives rise to an
associative multiplication in $\Lambda$ is guaranteed by the underlying
isomorphism of character rings
(see~\cite{fauser:jarvis:king:wybourne:2005a,fauser:jarvis:king:2007c}
for a direct proof). Associativity for such a deformed product is
equivalent to the 2-chain affiliated to $r_\pi$ being a 2-cyle, in the
appropriate Hopf algebra homology. For completeness, we give a short
derivation of this equivalence.
\mybenv{Lemma}\textbf{The 2-chain associated to $r_\pi$ is a 2-cycle}:\\
\label{lem:2cocyclerpi}
\myeenv
\noindent
\textbf{Proof}: We shall not need here the full theory of Hopf algebra
deformations, for which we refer 
to~\cite{sweedler:1968a,rota:stein:1994a,rota:stein:1994b,%
brouder:fauser:frabetti:oeckl:2002a,fauser:jarvis:2003a,%
fauser:jarvis:king:wybourne:2005a,fauser:jarvis:king:2007c} 
for details. The lemma is dual to the same statement for 2-cochains and
2-cocycles, and we proof the typographical simpler dual result. For any
3 symmetric functions $(f,g,h)$, the 2-cocycle condition for some
2-cochain $c: \Lambda \ot \Lambda \rightarrow {\mathbb C}$, reads in
terms of Sweedler parts
\begin{align}
  \sum c(g_{(1)},h_{(1)}) c(f,g_{(2)}h_{(2)}) 
  &= 
  \sum c(f_{(1)},g_{(1)}) c(f_{(2)}g_{(2)},h).
\end{align}
In the present case, affiliated with the 2-chain $r_\pi$ is the
following 2-cochain (here denoted simply $r$), 
\begin{align}
  r(f,g) &:= 
    \sum_\alpha \la f \vert \alpha^\pi_{(1)}\ra 
                \la g \vert \alpha^\pi_{(2)} \ra.
\end{align}
The circle product, whose associativity is crucial to the
$\pi$-Newell-Littlewood theorem, then reads
\begin{align}
  f\piprod g
  &= \sum r(f_{(1)},g_{(1)}) f_{(2)}g_{(2)}.
\end{align}
The expansion of 
$(f\piprod g)\piprod h = f\piprod (g\piprod h)$ in Sweedler parts yields
\begin{align}
  \sum r(f_{(1)},g_{(1)}) r\big( f_{(21)}g_{(21)}, h_{(1)}\big)  
     f_{(22)}g_{(22)}h_{(2)}
  &= \nn
  \sum r(g_{(1)},h_{(1)}) r\big( f_{(1)}, g_{(21)}h_{(21)}\big) 
  &  f_{(2)}g_{(22)}h_{(22)}.
\end{align}
However, using coassociativity and relabelling the Sweedler sums, this becomes
\begin{align}
  \sum r(f_{(11)},g_{(11)}) r(  f_{(12)}g_{(12)}, h_{(1)})  
    f_{(2)}g_{(2)}h_{(2)}
  &=\nn
  \sum r(g_{(11)},h_{(11)}) r( f_{(1)}, g_{(12)}h_{(12)}) 
  & f_{(2)}g_{(2)}h_{(2)},
\end{align}
wherein the coefficients of $f_{(2)}g_{(2)}h_{(2)}$ terms on each side
agree, because of the 2-cocycle condition applied to the triple 
$(f_{(1)},g_{(1)},h_{(1)})$.
\qed

As well as deforming the product, we can introduce an associated deformation of
the coproduct by dualising,
\begin{align}
  \Delta_\pi(f)
  &= 
  \sum_\alpha f_{(1)}\cdot\prod_{k=1}^p \alpha_k[\pi'_{(1)}] \ot 
       f_{(2)}\cdot \prod_{k=1}^p \alpha_k[\pi'_{(2)}] 
  \equiv R_\pi \comp \CO(f)
\end{align}
which is as indicated the composition of the standard outer coproduct with a 2-2
tangle $R_\pi$,
\begin{align}
  R_\pi &\cong\hskip-2ex 
  \vcenter{\hsize=0.12\textwidth\input{tangles/matRpi.tex}} 
  &&\mbox{where}
  & 
  R_{\pi}(f \ot g) 
    = \sum_\alpha f \cdot\prod_{k=1}^p \alpha_k[\pi'_{(1)}] \ot 
           g \cdot \prod_{k=1}^p \alpha_k[\pi'_{(2)}].
\end{align}
We have the following
\mybenv{Theorem}\label{thm:Quasitriangularityrpi}%
\textbf{Co-quasitriangularity of $\Lambda$ under $r_\pi$}:
The co-scalar product $r_\pi$ is a co-quasitriangular
structure~\cite{kassel:1995a} on the outer Hopf algebra $\Lambda$, namely it
fulfils the following properties
\begin{itemize}
\item[\textbf{(i)}] Normalization: 
     $(\varepsilon\otimes {\sf Id})\circ r_\pi 
     = \eta 
     = ({\sf Id}\otimes \varepsilon)\circ r_\pi$;
\item[\textbf{(ii)}] We have
\begin{align*}
  (a)\qquad
  \big( {\sf Id} \ot \CO \big) \,\comp\, r_\pi 
  &= r_\pi^{12}\cdot r_\pi^{13} \equiv ({\sf m} \ot {\sf Id}\ot {\sf Id})
     ({\sf Id} \ot {\sf sw} \ot {\sf Id})
     \,\comp\, r_\pi \ot r_\pi;
\\
  (b)\qquad
  \big( \CO \ot {\sf Id})\,\comp\, r_\pi 
  &= r_\pi^{13}\cdot r_\pi^{23} \equiv ({\sf Id}\ot {\sf Id}\ot {\sf m})
     ({\sf Id} \ot {\sf sw} \ot {\sf Id})
     \,\comp\, r_\pi \ot r_\pi.
\nonumber
\end{align*}
\item[\textbf{(iii)}] The antipode\footnote{%
        In $\Lambda$ we have $\antip^2={\sf Id}$ which simplifies the
        following relations.}
  relates $r_\pi$ and its convolutive inverse $r_\pi^{-1}$ as: 
  \begin{align*}
    (\antip\otimes {\sf Id})\circ r_\pi 
    &= 
       r_\pi^{-1}
    &&&    
    ({\sf Id}\otimes\antip)\circ r_\pi^{-1}
    &= 
       r_\pi
    \\    
    (\antip\otimes \antip)\circ r_\pi 
    &= 
       r_\pi
    &&&
    (\antip\otimes \antip)\circ r_\pi^{-1} 
    &= 
       r_\pi^{-1}
\end{align*}
\end{itemize}
\textbf{Proof}: For the normalization consider
\begin{align}
  (\varepsilon \ot {\sf Id}) \circ r_\pi
    = (\varepsilon \ot {\sf Id}) \sum (\alpha^\pi_{(1)}\ot \alpha^\pi_{(2)})
    = \sum \delta_{\alpha^\pi_{(1)},(0)}\, \alpha^\pi_{(2)}
    = \eta,
\end{align}
where the penultimate term has to be interpreted in the light
of~\eqref{eqn-willow}. The tangle diagrams for conditions
\textbf{(ii)}(a) and \textbf{(ii)}(b) read
\begin{align}
  \vcenter{\hsize=0.13\textwidth\input{tangles/rPi1213l.tex}}
  &\cong
  \vcenter{\hsize=0.13\textwidth\input{tangles/rPi1213r.tex}}
 &&&
  \vcenter{\hsize=0.13\textwidth\input{tangles/rPi1323l.tex}}
  &\cong
  \vcenter{\hsize=0.13\textwidth\input{tangles/rPi1323r.tex}}
\end{align}
To check that $r_\pi^{-1}$  as given in the first part of 
\textbf{(iii)} is the inverse of $r_\pi$ it suffices to note that
\begin{align}
  r_\pi^{-1} r_\pi 
    &=\sum_{\alpha,\beta} \alpha_{(1)}^\pi \beta_{(1)}^\pi \ot
        \antip(\alpha_{(2)}^\pi) \beta_{(2)}^\pi
     = \Delta^\prime(L_\pi) \Delta^\prime(M_\pi)
     = 1 = 1\ot 1 .
\end{align} 
The other cases are similar.
\qed

\mybenv{Corollary}\label{cor:RpiBraiding}\textbf{$\Lambda$ under $r_\pi$}:
The outer Hopf algebra of symmetric functions $\Lambda$ is a cobraided Hopf
algebra. Dually, $\Lambda$ with $R_\pi$ as defined above is a braided Hopf
algebra. $R_\pi$ satisfies the Yang-Baxter relation
\begin{align}
  R_\pi^{12} R_\pi^{13} R_\pi^{23}
  &= 
  R_\pi^{23} R_\pi^{13} R_\pi^{12} .
\end{align}
\myeenv

\noindent\textbf{Proof}: 
This follows directly from Kassel \cite{kassel:1995a} where the duality
between quasitriangular and coquasitriangular structures is established. To
emphasize the structure of the operators involved, however, we offer here a
direct proof that the object ${\texttt c}^\pi:= {\sf sw}\comp R_\pi$ is a
braid. To shorten the proof we again adopt the short-hand notation
$r_\pi =  \sum \alpha^\pi_{(1)} \ot \alpha^\pi_{(2)}$ for the Sweedler sums
occurring in the coscalar product, as well as
Theorem~\ref{thm:piNewellLittlewood} and $R_\pi$. Then we have
\begin{align}
  {\texttt c}^\pi(\lambda \otimes \mu)
  &= 
  \sum_\alpha \mu\cdot \prod_{k=1}^p\alpha_k{[}\pi'_{(1)}{]} \otimes 
    \lambda\cdot\prod_{k=1}^p\alpha_k{[}\pi'_{(2)}{]}
  := \sum_\alpha \mu\cdot\alpha^\pi_{(1)} \otimes 
      \lambda\cdot\alpha^\pi_{(2)}
\end{align}
We compute the left hand side of the braid equation (schematically
written $\texttt{c}^{12}\texttt{c}^{23}\texttt{c}^{12}
        =\texttt{c}^{23}\texttt{c}^{12}\texttt{c}^{23}$) as
\begin{align}
({\texttt c}^\pi \otimes {\sf Id})({\sf Id} \otimes {\texttt c}^\pi)({\texttt c}^\pi \otimes {\sf Id})(\lambda \otimes \mu \otimes \nu)
  &= \, ({\texttt c}^\pi \otimes {\sf Id})({\sf Id} \otimes {\texttt c}^\pi) (\mu\cdot\alpha^\pi_{(1)} \otimes \lambda\cdot\alpha^\pi_{(2)} \otimes \nu)\nonumber \\
  &= \,({\texttt c}^\pi \otimes {\sf Id})
   (\mu\cdot\alpha^\pi_{(1)} \otimes \nu\cdot\beta^\pi_{(1)} \otimes 
    \lambda\cdot\alpha^\pi_{(2)}\cdot\beta^\pi_{(2)} ) \nonumber \\
  &= \,(\nu\cdot\beta^\pi_{(1)}\cdot\gamma^\pi_{(1)} \otimes
    \mu\cdot\alpha^\pi_{(1)}\cdot\gamma^\pi_{(2)} \otimes 
    \lambda\cdot\alpha^\pi_{(2)}\cdot\beta^\pi_{(2)} ); \nonumber 
\end{align}    
the the right hand side is treated similarly,
\begin{align}
({\sf Id} \otimes {\texttt c}^\pi)({\texttt c}^\pi \otimes {\sf Id})({\sf Id} \otimes {\texttt c}^\pi)(\lambda \otimes \mu \otimes \nu)
  &= \,({\sf Id} \otimes {\texttt c}^\pi)({\texttt c}^\pi \otimes {\sf Id})
   (\lambda \otimes \nu\cdot\alpha^\pi_{(1)} \otimes \mu\cdot\alpha^\pi_{(2)} ) 
  \nn
  &= \,({\sf Id} \otimes {\texttt c}^\pi)
   (\nu\cdot\alpha^\pi_{(1)}\cdot\beta^\pi_{(1)} \otimes 
    \lambda\cdot\beta^\pi_{(2)} \otimes \mu\cdot\alpha^\pi_{(2)} )
  \nn
  &= \,(\nu\cdot\alpha^\pi_{(1)}\cdot\beta^\pi_{(1)} \otimes
    \mu\cdot\alpha^\pi_{(2)}\cdot\gamma^\pi_{(1)} \otimes  
    \lambda\cdot\beta^\pi_{(2)}\cdot\gamma^\pi_{(2)} ). \nonumber 
\end{align}
Then with the re-labelling of the summations to interchange $\alpha^\pi$ and 
$\gamma^\pi$, we have the required equality.
\qed

\section{Knots and links}\label{sec:KnotsnLinks}

\subsection{Knot alphabet}\label{subsec:KnotAlphabet}
In the introduction it was claimed that diagrammatic moves (tangle
diagrams) in the character ring $\Lambda$, referred to as the
\grpGL-alphabet \eqref{eq:GLalphabet}, could be used to assemble the
ingredients for describing isotopy classes of knots via their
two-dimensional projections, via a knot alphabet. In this section we
develop this relationship in detail, using the additional operations
on $\Lambda$ afforded by its realisation as a formal character ring
\CharHpi. As we shall see, the \grpGL-moves become a degenerate case
of the more general situation in \CharHpi. To evaluate the \CharHpi
tangle we \emph{rewrite} \CharHpi \emph{letters} into \CharGL \emph{words}. 
We shall refer to this as the transcription of the knot alphabet in
terms of the \grpH$_\pi$-alphabet. It is of paramount importance not to
confuse the trivial, in the sense of knot theory, \grpGL-tangles with
the \emph{nontrivial} transcription of the \CharHpi tangles into \CharGL
tangles. This gives us the following scheme (for the rewriting of crossings
and similarly for caps, cups)
\begin{align}\label{pic-plan}
\vcenter{\xymatrix@R1cm@C2cm{
   \mbox{~\hskip0.84cm}\mathfrak{B}_n~:~\raisebox{-2ex}{\mbox{\input{tangles/cDD.tex}}}
   \ar@{->}[r]^{\mbox{\tiny trivialize}}
   \ar@<5ex>@{->}[d]_{\mbox{\tiny hom}}
  &
   \mathfrak{S}_n~:~\raisebox{-2ex}{\mbox{
\scalebox{0.6} 
{
\begin{pspicture}(0,-1.025)(1.825,1.025)
\psline[linewidth=0.05cm](1.8,1.0)(0.0,-1.0)
\psline[linewidth=0.05cm](0.0,1.0)(1.8,-1.0)
\end{pspicture} 
}

}}
  \\
   \CharHpi~:~\raisebox{-2ex}{\mbox{\input{tangles/cDD.tex}}}
   \ar@{<->}[r]^{\mbox{\tiny rewrite (iso)}}
  &
   \grpGL~:~\raisebox{-2ex}{\mbox{\input{tangles/cDD2GL.tex}}}
   \ar@<-2.8ex>@{->}[u]_{\mbox{\tiny trivialize ($r_\pi=\eta\ot\eta$)}}
}}
\end{align}
We now introduce the required ingredients step by step, via the tangle
diagram and algebraic rewriting equivalents, defining the rewriting
isomorphism. As mentioned, a major role is played by the 2-2 tangle
crossing (braid) ${\texttt c}^\pi$ resulting from the \grpGL-\grpH$_\pi$
branching rule described in \S \ref{sec:CharHpi} above, and the
left-right and right-left dualities, represented by caps and cups.

The main structural change is a passage to \emph{oriented} tangles given
by a formal distinction between $\Lambda$ and its dual $\Lambda^*$. These
are of course isomorphic as linear spaces and as Hopf algebras via the
Schur-Hall scalar product, but the suite of moves entailed in the full
${\sf H}_\pi$-alphabet necessitates systematically accounting for
combinations of operators on tensor products of spaces of both types.
In fact this amounts to allowing co- and contravariant tensor characters
and rational representations. As a matter of convention we define the
following elementary 0-2 and 2-0 oriented tangles for the right-left
duality:
\begin{align}
  \bcap
  &\cong 
  \vcenter{\hsize=0.13\textwidth
\scalebox{0.6} 
{
\begin{pspicture}(0,-1.025)(2.025,1.025)
\psarc[linewidth=0.05](1.0,0.0){1.0}{-0.0}{180.0}
\psline[linewidth=0.05cm](2.0,0.0)(2.0,-1.0)
\psline[linewidth=0.04cm,arrowsize=0.05291667cm 3.5,arrowlength=1.5,arrowinset=0.4]{->}(1.1,1.0)(0.9,1.0)
\psline[linewidth=0.05cm](0.0,0.0)(0.0,-1.0)
\end{pspicture} 
}

};
  &&&
  \dcup 
  &\cong 
  \vcenter{\hsize=0.13\textwidth
\scalebox{0.6} 
{
\begin{pspicture}(0,-1.025)(2.025,1.025)
\psarc[linewidth=0.05](1.0,0.0){1.0}{180.0}{0.0}
\psline[linewidth=0.05cm](2.0,1.0)(2.0,0.0)
\psline[linewidth=0.04cm,arrowsize=0.05291667cm 3.5,arrowlength=1.5,arrowinset=0.4]{->}(1.1,-1.0)(0.8,-1.0)
\psline[linewidth=0.05cm](0.0,1.0)(0.0,0.0)
\end{pspicture} 
}

};
\nn
  1 
  &\mapsto
  \sum_\sigma \sigma \otimes \sigma^*; 
  &&&
  \lambda^* \otimes \mu  
  &\mapsto  
  \quad\langle \lambda | \mu\rangle , \nonumber
\end{align}
whose compositions implement the topological move $\textbf{R0}$ on $\Lambda$
and $\Lambda^*$
\begin{align}
   \vcenter{\hsize=0.13\textwidth\input{tangles/reidemeister0ldown.tex}}
   &\cong
   \vcenter{\hsize=0.05\textwidth
\scalebox{0.6} 
{
\begin{pspicture}(0,-1.02)(0.02,1.02)
\psline[linewidth=0.04cm](0.0,1.0)(0.0,-1.0)
\psline[linewidth=0.04cm,arrowsize=0.05291667cm 3.0,arrowlength=1.5,arrowinset=0.4]{->}(0.0,0.2)(0.0,-0.1)
\end{pspicture} 
}

},
  &&&
   \vcenter{\hsize=0.05\textwidth
\scalebox{0.6} 
{
\begin{pspicture}(0,-1.02)(0.02,1.02)
\psline[linewidth=0.04cm](0.0,1.0)(0.0,-1.0)
\psline[linewidth=0.04cm,arrowsize=0.05291667cm 3.0,arrowlength=1.5,arrowinset=0.4]{->}(0.0,-0.1)(0.0,0.2)
\end{pspicture} 
}

}
   &\cong
   \vcenter{\hsize=0.13\textwidth\input{tangles/reidemeister0rup.tex}}.
\end{align}
Next we turn to the major ingredients of the knot alphabet, the braid ${\texttt
c}^\pi$ and its inverse $({\texttt c}^\pi)^{-1}$. For these, the following
over-under crossing representations are adopted to stand for the indicated
tangle diagrams; the repositioning of $r_\pi$ (for example as compared with the
tangle diagram for $R_\pi$ above) is easily checked:
\mybenv{Definition}\label{lemma:braiddiags}
\textbf{${\sf H}_\pi$ alphabet structure of ${\texttt c}^\pi$ and $({\texttt c}^\pi)^{-1}$}: \\
We have
\begin{align}  
  \left.\vcenter{\hsize=0.11\textwidth\input{tangles/cDD.tex}}
  \right\vert_{\textsf{H}_\pi}
   &\cong
  \left.\vcenter{\hsize=0.15\textwidth\input{tangles/cDDgl.tex}}
  \right\vert_{\textsf{GL}}
 &&&
  \left.\vcenter{\hsize=0.11\textwidth\input{tangles/ciDD.tex}}
  \right\vert_{\textsf{H}_\pi}
  &\cong
  \left.\vcenter{\hsize=0.15\textwidth\input{tangles/ciDDgl.tex}}
  \right\vert_{\textsf{GL}}
\end{align}
where $r_\pi^{-1}$ is the inverse of $r_\pi$, namely 
(see Theorem~\ref{thm:Quasitriangularityrpi}),
\begin{align}
  r_\pi 
  &=
     \sum \alpha^\pi_{(1)} \ot \alpha^\pi_{(2)}, \qquad
  r_\pi^{-1} 
   = \sum {\sf S}(\alpha^\pi_{(1)}) \ot \alpha^\pi_{(2)}
\end{align}
\myeenv
Still missing from the compendium are the companion dual cap and cup diagrams
$\bpcap$ and $\dpcup$:
\begin{align}
  \bpcap 
  &\cong 
  \vcenter{\hsize=0.13\textwidth
\scalebox{0.6} 
{
\begin{pspicture}(0,-1.025)(2.025,1.025)
\psarc[linewidth=0.05](1.0,0.0){1.0}{-0.0}{180.0}
\psline[linewidth=0.05cm](2.0,0.0)(2.0,-1.0)
\psline[linewidth=0.04cm,arrowsize=0.05291667cm 3.5,arrowlength=1.5,arrowinset=0.4]{->}(0.9,1.0)(1.1,1.0)
\psline[linewidth=0.05cm](0.0,0.0)(0.0,-1.0)
\end{pspicture} 
}

};
  &&&  
  \dpcup
  &\cong 
  \vcenter{\hsize=0.13\textwidth
\scalebox{0.6} 
{
\begin{pspicture}(0,-1.025)(2.025,1.025)
\psarc[linewidth=0.05](1.0,0.0){1.0}{180.0}{0.0}
\psline[linewidth=0.05cm](2.0,1.0)(2.0,0.0)
\psline[linewidth=0.04cm,arrowsize=0.05291667cm 3.5,arrowlength=1.5,arrowinset=0.4]{->}(1.0,-1.0)(1.1,-1.0)
\psline[linewidth=0.05cm](0.0,1.0)(0.0,0.0)
\end{pspicture} 
}

};
\end{align}
whose compositions implement the corresponding straightening rules
\begin{align}
   \vcenter{\hsize=0.13\textwidth\input{tangles/reidemeister0lup.tex}}
   &\cong
   \vcenter{\hsize=0.05\textwidth
\scalebox{0.6} 
{
\begin{pspicture}(0,-1.02)(0.02,1.02)
\psline[linewidth=0.04cm](0.0,1.0)(0.0,-1.0)
\psline[linewidth=0.04cm,arrowsize=0.05291667cm 3.0,arrowlength=1.5,arrowinset=0.4]{->}(0.0,-0.1)(0.0,0.2)
\end{pspicture} 
}

},
  &&&
   \vcenter{\hsize=0.05\textwidth
\scalebox{0.6} 
{
\begin{pspicture}(0,-1.02)(0.02,1.02)
\psline[linewidth=0.04cm](0.0,1.0)(0.0,-1.0)
\psline[linewidth=0.04cm,arrowsize=0.05291667cm 3.0,arrowlength=1.5,arrowinset=0.4]{->}(0.0,0.2)(0.0,-0.1)
\end{pspicture} 
}

}
   &\cong
   \vcenter{\hsize=0.13\textwidth\input{tangles/reidemeister0rdown.tex}}
\end{align}
As we now show, the $\bpcap$, $\dpcup$ moves must be \emph{derived} from the
foregoing ingredients and cannot be independently defined. Firstly, note that
the following diagram representing an elementary twist, or formally,
\emph{writhe}, is represented by a linear operator which will play a crucial
role:
\mybenv{Definiton}\label{def:writhe}\textbf{Writhe operator}:
\begin{align}
  \vcenter{\hsize=0.1\textwidth\input{tangles/writheRight.tex}}
  &\Leftrightarrow \quad
 \lambda \mapsto \theta_\pi (\lambda).
\end{align}
\myeenv
The existence of the writhe operator $\theta_\pi$ signifies that the knot
diagrammar which we are constructing from the ${\sf H}_\pi$-alphabet describes
knot projections in ambient isotopy, that is including \emph{framing} as part of
the crossing information. Indeed, the removal of the twist represented by
$\theta_\pi$ is given by its inverse, implementing the modified (Reidemeister
\textbf{R1}$^\prime$) move~\eqref{reidemeisterI}, 
\begin{align}
  \vcenter{\hsize=0.1\textwidth\input{tangles/reidemeister1DDright.tex}}
  &\cong
   \vcenter{\hsize=0.06\textwidth
\scalebox{0.6} 
{
\begin{pspicture}(0,-1.02)(0.02,1.02)
\psline[linewidth=0.04cm](0.0,1.0)(0.0,-1.0)
\psline[linewidth=0.04cm,arrowsize=0.05291667cm 3.0,arrowlength=1.5,arrowinset=0.4]{->}(0.0,0.2)(0.0,-0.1)
\end{pspicture} 
}

}
   \cong
  \vcenter{\hsize=0.1\textwidth\input{tangles/reidemeister1UUright.tex}}
\end{align}
Introducing a non-trivial writhe entails in the algebraic setting working
with ribbon Hopf algebras.
 
We proceed to the dual cup and cap. We can rewrite these composed with
untwisting moves, and rearranging, using only the assumed straightening
operations for $\bcap$, $\dcup$, $\bpcap$ and $\dpcup$, in order to derive
alternative forms:
\mybenv{Lemma}\label{lemma:DerivedCupCap}
\textbf{Dual cap $\bpcap$ and cup $\dpcup$}: we have 
\begin{align}\label{eq:bpdpDef}
  \bpcap 
  &:= \big(  {\sf Id}_\Lambda \ot (\theta_\pi)^{-1} \big) 
      \comp {\texttt c}^\pi_{\Lambda \Lambda^*} 
      \comp \bcap, \qquad 
  &&&
  \dpcup 
  &:= \dcup 
      \comp \overline{\texttt c}^\pi_{\Lambda \Lambda^*} 
      \comp \big( (\theta_\pi)^{-1}\ot {\sf Id}_{\Lambda^*} \big).
\end{align}
\mbox{}\myeenv
\noindent\textbf{Proof}:
Observe the equivalence of $\bpcap$ and $\dpcup$ to the following respective
tangle diagrams which feature the original 0-2 and 2-0 tangles $\bcap$ and
$\dcup$, and evaluate:
\begin{align}
  \bpcap :
  \vcenter{\hsize=0.12\textwidth
\scalebox{0.6} 
{
\begin{pspicture}(0,-1.025)(2.025,1.025)
\psarc[linewidth=0.05](1.0,0.0){1.0}{-0.0}{180.0}
\psline[linewidth=0.05cm](2.0,0.0)(2.0,-1.0)
\psline[linewidth=0.04cm,arrowsize=0.05291667cm 3.5,arrowlength=1.5,arrowinset=0.4]{->}(0.9,1.0)(1.1,1.0)
\psline[linewidth=0.05cm](0.0,0.0)(0.0,-1.0)
\end{pspicture} 
}

}
  &:= 
  \vcenter{\hsize=0.12\textwidth\input{tangles/capRightWrithe.tex}}
  &&&
  \dpcup :
  \vcenter{\hsize=0.12\textwidth
\scalebox{0.6} 
{
\begin{pspicture}(0,-1.025)(2.025,1.025)
\psarc[linewidth=0.05](1.0,0.0){1.0}{180.0}{0.0}
\psline[linewidth=0.05cm](2.0,1.0)(2.0,0.0)
\psline[linewidth=0.04cm,arrowsize=0.05291667cm 3.5,arrowlength=1.5,arrowinset=0.4]{->}(1.0,-1.0)(1.1,-1.0)
\psline[linewidth=0.05cm](0.0,1.0)(0.0,0.0)
\end{pspicture} 
}

}
  &:= 
  \vcenter{\hsize=0.12\textwidth\input{tangles/cupRightWrithe.tex}} .
\end{align}
\qed

The price paid for these rearrangements is the introduction of additional
auxiliary crossings, related to ${\texttt c}^\pi$, which we now explain.
Evidently, braidings on different spaces require labels such as ${\texttt
c}_{V,W}$. Adopting this notation, the original braid ${\texttt c}^\pi$ is here
denoted ${\texttt c}^\pi_{\Lambda, \Lambda}$ or  ${\texttt c}^\pi_{\Lambda
\Lambda}$ for simplicity.  Denoting right chirality by an overbar, for a choice
of spaces $\Lambda$, $\Lambda^*$ on each line there are thus $2^3=8$ types,
namely ${\texttt c}^\pi_{\Lambda \Lambda}$ and $\overline{\texttt
c}{}^\pi_{\Lambda \Lambda}$ (the inverse $({\texttt c}^\pi_{\Lambda
\Lambda})^{-1}$, as derived above), together with  ${\texttt c}^\pi_{\Lambda^*
\Lambda}$ and $\overline{\texttt c}{}^\pi_{\Lambda^* \Lambda}$, ${\texttt
c}^\pi_{\Lambda^* \Lambda^*}$ and $\overline{\texttt c}^\pi_{\Lambda^*
\Lambda^*}$, as well as ${\texttt c}^\pi_{\Lambda \Lambda^*}$ and
$\overline{\texttt c}^\pi_{\Lambda \Lambda^*}$. 

Dualisation and straightening of the basic crossings, together with appropriate
identification of inverses, suffice for evaluation of all cases (including those
needed to define $\bpcap$ and $\dpcup$). The following lemma provides a complete
dictionary of these crossings (including, for completeness, the standard ones),
both in diagrammatic form, and as well as giving their explicit action on
elements of the appropriate space $\Lambda^{(*)} \otimes \Lambda^{(*)}$, in the
$S$-function basis:
\mybenv{Lemma}\label{lemma:AuxiliaryCrossings}
\textbf{Knot to \CharHpi dictionary -- I. Crossings}:
In addition to the standard crossings listed, the table~\ref{table-crossings}
defines the auxiliary crossings in terms of oriented over- and under-crossings,
in terms of primitive tangles involving operations in \CharGL, and in terms
of their action on characters belonging to the relevant spaces,
see table~\ref{table-crossings}.
\begin{table}
\caption{Crossings, \CharHpi, \CharGL tangles and algebraic forms of the
         8 possible forms of oriented crossings.\label{table-crossings}}
\begin{tabular}[tfbp]{|c|c|c|rcl|}
\hline
 &&&&&\\[-1ex]  braid & \CharHpi\ tangle &\CharGL\ tangle
      &\multicolumn{3}{c|}{algebraic expression} \\
 &&&&&\\[-1ex]
\hline
 &&&&&\\[-1ex]
  $\texttt{c}^\pi_{\Lambda \Lambda}$ 
  & 
  $\vcenter{\hsize=0.15\textwidth\input{tangles/cDD2.tex}}$
  & 
  $\vcenter{\hsize=0.15\textwidth\input{tangles/cDD2GL.tex}}$
  & 
   $\lambda \otimes \mu$ & $\mapsto$ & 
   $\sum_\alpha \mu\!\cdot\!\alpha^{\pi}_{(2)} \otimes \lambda\!\cdot\!\alpha^{\pi}_{(1)}$
\\ &&&&&\\[-1ex]
  $\overline{\texttt{c}}^\pi_{\Lambda \Lambda}$ 
  & 
  $\vcenter{\hsize=0.15\textwidth\input{tangles/ciDD2.tex}}$
  & 
  $\vcenter{\hsize=0.15\textwidth\input{tangles/ciDD2GL.tex}}$
  & 
   $\lambda \otimes \mu$ & $\mapsto$ & 
   $\sum_\alpha \mu\!\cdot\!{\sf S}(\alpha^{\pi}_{(2)}) \otimes 
    \lambda\!\cdot\!\alpha^{\pi}_{(1)}$
\\
  &&&&&\\[-1ex]
\hline  
  &&&&&
\\[-1ex]
  $\texttt{c}^\pi_{\Lambda^* \Lambda}$
  & 
  $\vcenter{\hsize=0.16\textwidth\input{tangles/cUD2.tex}}$
  & 
  $\vcenter{\hsize=0.15\textwidth\input{tangles/cUD2GL.tex}}$
  & 
   $\lambda^* \otimes \mu$ & $\mapsto$ & 
   $\sum_\alpha \mu\!\cdot\! \alpha^\pi_{(2)} \otimes
   ( \lambda\! / \!\alpha^\pi_{(1)})^*$ 
\\ &&&&&\\[-1ex]
  $\overline{\texttt{c}}^\pi_{\Lambda^* \Lambda}$ 
  & 
  $\vcenter{\hsize=0.16\textwidth\input{tangles/ciUD2.tex}}$
  & 
  $\vcenter{\hsize=0.15\textwidth\input{tangles/ciUD2GL.tex}}$
  & 
   $\lambda^* \otimes \mu$ & $\mapsto$ & 
   $\sum_\alpha \mu\!\cdot\! {\sf S}(\alpha^\pi_{(2)}) \otimes 
    (\lambda\! / \!\alpha^\pi_{(1)})^*$ 
\\ 
  &&&&&\\[-1ex]
\hline
  &&&&&
\\[-1ex]
  $\texttt{c}^\pi_{\Lambda^* \Lambda^*}$
  & 
  $\vcenter{\hsize=0.16\textwidth\input{tangles/cUU2.tex}}$
  & 
  $\vcenter{\hsize=0.16\textwidth\input{tangles/cUU2GL.tex}}$
  & 
   $\lambda^* \otimes \mu^*$ & $\mapsto$ & 
   $\sum_\alpha (\mu\! / \!\alpha^\pi_{(2)})^* \otimes (\lambda\! / \!\alpha^\pi_{(1)})^*$
\\ &&&&&\\[-1ex]
  $\overline{\texttt{c}}^\pi_{\Lambda^* \Lambda^*}$
  & 
  $\vcenter{\hsize=0.16\textwidth\input{tangles/ciUU2.tex}}$
  & 
  $\vcenter{\hsize=0.16\textwidth\input{tangles/ciUU2GL.tex}}$
  & 
   $\lambda^* \otimes \mu^*$ & $\mapsto$ & 
   $\sum_\alpha (\mu\! / \!{\sf S}(\alpha^\pi_{(2)}))^* \otimes 
    (\lambda\! / \!\alpha^\pi_{(1)})^*$ 
\\
  &&&&&\\[-1ex]
\hline
  &&&&&
\\[-1ex]
  $\texttt{c}^\pi_{\Lambda \Lambda^*}$
  & 
  $\vcenter{\hsize=0.16\textwidth\input{tangles/cDU2.tex}}$
  &
  $\vcenter{\hsize=0.16\textwidth\input{tangles/cDU2GL.tex}}$
  & 
   $\lambda \otimes \mu^*$ & $\mapsto$ & 
   $\sum_\alpha (\mu\! / \!\alpha^\pi_{(2)})^* \otimes 
    \lambda \!\cdot\!\alpha^\pi_{(1)}$ 
\\ &&&&&\\[-1ex]
  $\overline{\texttt{c}}^\pi_{\Lambda \Lambda^*}$
  & 
  $\vcenter{\hsize=0.16\textwidth\input{tangles/ciDU2.tex}}$
  &
  $\vcenter{\hsize=0.16\textwidth\input{tangles/ciDU2GL.tex}}$
  & 
   $\lambda \otimes \mu^*$ & $\mapsto$ & 
   $\sum_\alpha (\mu \! / \!{\sf S}(\alpha^\pi_{(2)}))^* \otimes 
    \lambda\!\cdot\!\alpha^\pi_{(1)}$   
\\ &&&&&\\
\hline
\end{tabular}
\end{table}
\mbox{}\myeenv

\noindent\textbf{Proof}:
In table~\ref{table-crossings} we construct the six auxiliary crossings
for partly upward oriented lines representing elements of $\Lambda^*$,
and we provide the rewriting homomorphism \CharHpi to \CharGL for all crossings.
The left column gives the name of the crossing, the \CharHpi column
provides the tangle version including orientation, and further shows
how the auxiliary crossing is obtained by using the caps and cups.
Hence the crossing appears as either ${\texttt c}^\pi_{U V}$ or 
$\overline{\texttt c}^\pi_{U V}$ with $U,V\in \{\Lambda,\Lambda^*\}$. 
The first four auxiliary crossings use composition with $\bcap$ and
$\dcup$. The last two cannot be obtained this way, as we would need
$\overline{\bcap}_\pi$ and $\overline{\dcup}_\pi$ tangles, which we
define from the $\bcap,\dcup$ closed structure and the crossing.
However, these two cases can be obtained as inverse tangles under
vertical composition with
$\overline{\texttt{c}}^\pi_{\Lambda^*\Lambda}$ and
$\texttt{c}^\pi_{\Lambda^*\Lambda}$. The fact that we can produce
inverses under both vertical and horizontal composition relies upon
the identity $r_\pi^{-1}r_\pi = 1$. The \CharGL tangles provide the
image of the crossings in \CharGL\kern-0.6ex, and hence the \grpGL-word which
faithfully represents the crossing. The last column gives the
algebraic expression of the crossings.

For a proof we present the algebraic identity, showing how the 
switch is modified by the insertion of terms $\alpha^\pi$ or
${\sf S}(\alpha^\pi)$ coming from the crossing. The main thing to note
is that on $\Lambda$ one acts by Schur function multiplication, while
on $\Lambda^*$ one acts by skewing. Following the fate of such summands
in the sliced tangle diagram for one case, we have
\begin{align}
{\texttt c}^\pi_{\Lambda^* \Lambda}:  \lambda^* \ot \mu 
  &\mapsto \, 
    \sum_\sigma \lambda^* \ot \mu \ot \sigma \ot \sigma^* 
  \nn
  &\mapsto \, 
    \sum_{\sigma,\alpha} \lambda^* \ot \sigma \!\cdot\! \alpha^\pi_{(1)} \ot \mu \!\cdot\! \alpha^\pi_{(2)}\ot \sigma^* 
    \nn
  & \mapsto \,    
    \sum_{\sigma,\alpha} \la \lambda\vert  \sigma \!\cdot\! \alpha^\pi_{(1)} \ra \mu \!\cdot\! \alpha^\pi_{(2)}\ot \sigma^* 
   \nn
   & = \,
    \sum_{\sigma,\alpha} \la \lambda\!/\!\alpha^\pi_{(1)} \vert  \sigma \ra  \mu \!\cdot\!\alpha^\pi_{(2)}\ot \sigma^* 
   \nn
   & = \,
    \sum_{\alpha} \mu\!\cdot\! \alpha^\pi_{(2)} \ot ( \lambda\!/\!\alpha^\pi_{(1)})^*   
\end{align}
with the sum over $\sigma$ enforcing $\lambda^* \mapsto (\lambda\!/\!\alpha^\pi)^*$
on each dual line. In this way, the Sweedler parts of $r_\pi$ involved in the
crossing systematically appear with outer skew rather than outer product when
they appear with elements of  $\Lambda^*$, while the crossing handedness is
still reflected in the presence of $(r_\pi)^{-1}$ via the antipode.
\qed

  With the above list of crossings in hand, we turn to the evaluation of
explicit forms for $\bpcap$, $\dpcup$ and
$\theta_\pi$ to complete the transcription of the knot alphabet in \CharHpi
into moves in \CharGL. We have:

\mybenv{Lemma}\label{lemma:Knot2HpiDictionary}
\textbf{Knot to \CharHpi dictionary -- II}: 
In addition to the basic crossings defined in
Lemma \ref{lemma:AuxiliaryCrossings} above, the following elements of
the \CharHpi-knot alphabet are defined diagrammatically and by their
action on states, see table \ref{table-cups}:
\begin{table}
\caption{left-right cups, caps, twists, as \CharHpi, \CharGL tangles
         and their algebraic forms\label{table-cups}}
\begin{tabular}[tfbp]{|c|c|c|rcl|}
\hline
 &&&&&\\[-1ex]  map & \CharHpi\ tangle &\CharGL\ tangle
      &\multicolumn{3}{c|}{algebraic expression} \\
 &&&&&\\[-1ex]
\hline
 &&&&&\\[-1ex]
  $\bcap$
  &
  $\vcenter{\hsize=0.2\textwidth\input{tangles/capLeft.tex}}$
  &
  $\vcenter{\hsize=0.12\textwidth\input{tangles/cap.tex}}$
  & 
    $1$ & $\mapsto$ & 
    $\sum \sigma \otimes \sigma^*$
\\
  &&&&&\\[-1ex]
  $\dcup$ 
  & 
  $\vcenter{\hsize=0.2\textwidth\input{tangles/cupLeft.tex}}$
  & 
  $\vcenter{\hsize=0.12\textwidth\input{tangles/cup.tex}}$
  &
   $\lambda^* \otimes \mu $ & $\mapsto$ & 
   $\langle \lambda | \mu \rangle  $ 
\\
  &&&&&\\[-1ex]
  $\bpcap$ 
  & 
  $\vcenter{\hsize=0.09\textwidth\input{tangles/capRight.tex}}$
  $\cong$
  $\vcenter{\hsize=0.12\textwidth\input{tangles/capRightWrithe.tex}}$
  & 
  $\vcenter{\hsize=0.12\textwidth\input{tangles/cap.tex}}$
  &
   $1$ & $\mapsto$ & 
   $\sum   \rho^*\otimes \rho$ 
\\
  &&&&&\\[-1ex]
  $\dpcup$
  &
  $\vcenter{\hsize=0.09\textwidth\input{tangles/cupRight.tex}}$
  $\cong$
  $\vcenter{\hsize=0.12\textwidth\input{tangles/cupRightWrithe.tex}}$
  & 
  $\vcenter{\hsize=0.12\textwidth\input{tangles/cup.tex}}$
  &
   $\lambda \otimes \mu^* $ & $\mapsto$ & 
   $\langle \mu | \lambda \rangle  $
\\
  &&&&&\\[-1ex]
  $\theta_\pi$ 
  &
  $\vcenter{\hsize=0.1\textwidth\input{tangles/writheInvLeft.tex}}\cong
   \vcenter{\hsize=0.1\textwidth\input{tangles/writheRight.tex}}$
  & 
  $\vcenter{\hsize=0.1\textwidth\input{tangles/tadpoleR.tex}}$
  &
   $\lambda$ & $\mapsto$ & 
   $Q_\pi \cdot \lambda  $
\\
  &&&&&\\[-1ex]
  $(\theta_\pi)^{-1}$
  &
  $\vcenter{\hsize=0.09\textwidth\input{tangles/writheInvRight.tex}}\cong
   \vcenter{\hsize=0.09\textwidth\input{tangles/writheLeft.tex}}$
  & 
  $\vcenter{\hsize=0.1\textwidth\input{tangles/tadpoleiR.tex}}$
  & 
  $\lambda$ & $\mapsto$ & 
  $(Q_\pi  )^{-1} \cdot \lambda$ 
\\[-1ex]
&&&&&\\
\hline
\end{tabular}
\end{table}
\mbox{}\myeenv

\noindent
\textbf{Proof}:\\
As already noted, the equivalence (lemma \ref{lemma:DerivedCupCap}) of $\bpcap$ and $\dpcup$ to tangle diagrams which feature the original 0-2 and 2-0 tangles $\bcap$ and $\dcup$ can now be exploited in the light of the explicit forms for the auxiliary crossings
given above (lemma \ref{lemma:AuxiliaryCrossings}). Taking for example the expression for $\bpcap$, we have
for its action in sliced form
\begin{align}
  1 \mapsto \sum \sigma \otimes \sigma^*   
  &\mapsto 
  \sum (\sigma/\alpha^\pi_{(1)})^* \ot 
        \sigma \cdot \alpha^\pi_{(2)} \nn
  &\mapsto 
  \sum (\sigma/\alpha^\pi_{(1)})^* \ot 
       (\theta_\pi)^{-1}\big(\sigma \cdot \alpha^\pi_{(2)} \big).
\end{align}
However as a consequence of outer product associativity, $\sum_\sigma
\sigma/\alpha \ot \sigma \cdot \beta = \sum_\sigma \sigma \ot \sigma \cdot
(\alpha \!\cdot \!\beta)$ and for the Sweedler sums entailed in $r_\pi$ this
just produces the Cauchy scalar element $Q_\pi$ on the right-hand side,
namely we infer
$\bpcap(1) = \sum \sigma^* \ot (\theta_\pi)^{-1}\big( Q_\pi \cdot \sigma \big)$.
On the other hand, the expression for $\bpcap$ could have been written using the
inverse Reidemeister move instead. Using this alternative, and following the
same progression through the sliced diagram now leads to $\bpcap(1) = \sum
\sigma^* \ot \theta_\pi\big( (Q_\pi)^{-1} \cdot \sigma\big)$. Equating these
requires $\theta_\pi\big( (Q_\pi)^{-1} \cdot \sigma\big) =
(\theta_\pi)^{-1}\big( Q_\pi \cdot \sigma \big)$ or $ (\theta_\pi)^{2}\big( \rho
\big) = (Q_\pi)^{2} \cdot \rho$ for arbitrary $\rho$. We solve this functional
equation by simply \emph{identifying}  the linear operator $\theta_\pi$ with the
multiplication by $Q_\pi$. This immediately means that $\bpcap$, and similarly
$\dpcup$, collapse to their minimal forms, with $\theta_\pi$ corresponding to
multiplication by $Q_\pi$, as claimed.

  Alternatively, the elementary twist itself can be analysed via the
algebraic steps given in the corresponding sliced tangle diagram of 
$\bpcap$, and we find a consistent result, namely
\begin{align}
  \vcenter{\hsize=0.1\textwidth\input{tangles/writheRight.tex}}
  &\Leftrightarrow  
  &&& 
  \lambda 
  &\mapsto 
  \sum \lambda \ot \sigma \ot \sigma^* 
\nn
  &&&&&\mapsto 
  \sum  \sigma \cdot \alpha^\pi_{(1)} \ot 
       \lambda \cdot \alpha^\pi_{(2)} \ot 
       \sigma^* 
\nn
  &&&&&\mapsto
  \sum \sigma \cdot \alpha^\pi_{(1)} \,
       \sigma^* (\lambda \cdot \alpha^\pi_{(2)}) 
  \equiv
  \sum  \alpha^\pi_{(1)} \cdot 
  \left( \sum_\sigma \sigma \, \langle \sigma | 
         \lambda \cdot \alpha^\pi_{(2)} \rangle \right) 
\nn
  &&&&&=
  \lambda \cdot \sum \big(\alpha^\pi_{(1)}  \cdot \alpha^\pi_{(2)}\big) 
  \equiv Q_\pi \cdot \lambda
\end{align}
as required. In the second and third lines the forms of the standard cap- and
dual cup- operators $\bcap$ and $\dpcup$ have been implemented, together with
the resolution of the identity as ${\sf Id} = \sum_\sigma \sigma \comp \sigma^*$
in an orthonormal basis). As a consistency check, if in lemma
\ref{lemma:AuxiliaryCrossings} above, the three crossings derived as inverses
are instead evaluated in terms of standard crossings, using `unstraightening'
moves involving $\bpcap$ and $\dpcup$, the same expressions result.

Finally we need to check that the alternative versions of the $\bpcap,\dpcup$
and twist tangles yield the same translation to ensure that our map is indeed
a homomorphism. This can be done by a tangle argument for all cases in the
table~\ref{table-cups}.
\qed

The last two lemmas \ref{lemma:AuxiliaryCrossings} and 
\ref{lemma:Knot2HpiDictionary} can be summarized as
\mybenv{Proposition}
The \emph{rewrite} map from the closed braid part of \CharHpi into
\CharGL given in~\ref{pic-plan} is an isomorphism of groups.
\myeenv
Equipped with this structure we finally obtain the
\mybenv{Theorem}\label{thm:MainTheorem}
\textbf{$\Lambda$ as a ribbon Hopf algebra}:
For each partition $\pi$ the space $\Lambda\cong \CharHpi$ equipped with the
braid operators ${\texttt c}^\pi$, $({\texttt c}^\pi){}^{-1}$ together with the
objects   $\bcap$, $\dcup$, $\bpcap$, $\dpcup$ and the canonical writhe element
$Q_\pi$, is a ribbon Hopf algebra \cite{kassel:1995a}.
\myeenv

\subsection{Knot invariant operators}
\label{subsec:KnotInvariants}
The identification of all the ingredients of the knot alphabet in terms of moves in \CharHpi confers on it a status equivalent to 
that of \CharGL itself. Note that the presentation of tangle relations in \CharGL (see Figure \ref{fig:GLrelations} following Theorem \ref{thm:OuterHopf}, \S \ref{sec:CharGL}), contains
several implicit steps such as the introduction of cup and cap operators and their duals. In the \CharHpi case, by contrast,
these have had to be considered in detail.  Although the final forms for \bcap, \dcup, $\bpcap$ and $\dpcup$ are indeed equivalent
to those for \CharGL (adapted to the case of directed tangles), it should be emphasised that these definitions are by no means automatic. 

We finally return to the 
discussion of the introduction to the paper, \S \ref{sec:Intro}, where it was mentioned that one aim of our current exploration of manipulations of goup characters rings was the possibility of generating new knot invariants.
With the full apparatus of ribbon Hopf algebras in place,
we can now apply the well-established formalism (see \S \ref{sec:Intro}) whereby tangle diagrams can be associated with knot projections, with their evaluations in $\Lambda$ and its tensor powers guaranteed to be invariants. 

Complete knots and links are projected as decorated images of products of circles, and so must be interpreted in terms of slicings of 0-0 tangles. Consider the oriented unknots, and their evaluation according to the obvious slicings; for example (in the left-handed case)
\begin{align}
  \vcenter{\hsize=0.3\textwidth\input{tangles/traceId.tex}}  
  &\Leftrightarrow
  &&& 
  \begin{array}{rcl}
    1 &\mapsto & \\
      &\mapsto &
      \sum_\sigma  \sigma \otimes \sigma^*  \\
      &\mapsto & 
      \sum_\sigma \langle \sigma | \sigma \rangle \equiv \sum_\sigma 1,
  \end{array}
\end{align}
a formally infinite sum. It is clear however that a finite result can be arranged for this and other links and knots by simply cutting and opening one strand (or  more) of the 0-0 tangle, and evaluating the resulting 1-1 tangle invariant as an element of $End(\Lambda)$ (or of $End(\otimes^k \Lambda)$ for a $k$-$k$ tangle). Clearly, for the left- and right-handed unknots, this simply produces the identity operator on $\Lambda$ or $\Lambda^*$, respectively, but is potentially more interesting for general knots and links. 

Consider then a general link projection, made into a 1-1 tangle as
described. By standard theorems it is regular-isotopic to the corresponding
single-strand opening of a braid word representation of the knot or link,
and it is in this canonical form that we proceed with an evaluation. Thus
suppose that the knot $K$ is isotopic to the closure of a braid element
in ${\mathfrak B}_m$, by means of a word of length $\ell$,
$b_K = b_{i_1}{}^{e_1} b_{i_2}{}^{e_2} \cdots b_{i_\ell}{}^{e_\ell}$, where
each $i_k \in {\{}1,2,\cdots,m\!-\!1{\}}$, with each exponent $e_i = \pm 1$. 
Clearly $\ell \equiv \sum_i |e_i|$, while the sum
$w = \sum_i e_i := w_+-w_-$ is the writhe of the knot or link
projection (the difference between the positive and negative exponent sums).
\begin{align}\label{fig:Knot8-1}
    \vcenter{\hsize=0.2\textwidth\input{tangles/knot8-1.tex}}
    &\Leftrightarrow
    \vcenter{\hsize=0.19\textwidth\input{tangles/braid8-1.tex}}
    &&
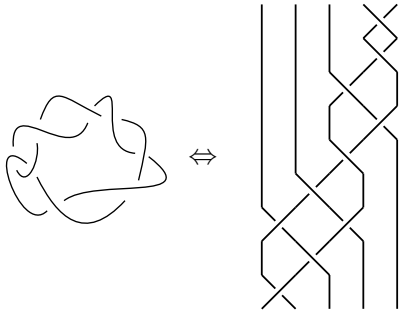
    \vcenter{\hsize=0.45\textwidth\noindent The knot \texttt{8\_1} and its braid
    representation.}
\end{align}
The situation is as illustrated in the figure \ref{fig:Knot8-1}, which
shows the specific case of the knot $\texttt{8\_1}$ and a braid
representation of it\footnote{Taken from the Rolfsen knot table,
\small{\texttt{http://katlas.org/wiki/The\_Rolfsen\_Knot\_Table}}}. 
Writhes are introduced under the map from the knot to the braid
provided by the Alexander theorem but they always appear in mutually
inverse pairs on the same line.

In order to write down the operator in $\Lambda$ depicted by the sliced
diagram for such a braid tangle, using the above-developed toolkit in
\CharHpi, it is only necessary to use the rules for the basic crossing
${\texttt c}_\pi$ and its inverse. Schematically, the representation of
the knot as a braid closure in ${\mathfrak B}_m$ means that there will
be a nested sequence of $m\!-\!1$ cap tangles, and corresponding cups,
together with one open line, say the first, representing the cut strand.
Assuming that the (downward) strands are labelled $\sigma_1, \sigma_2,
\cdots, \sigma_m$, with $\sigma_1$ the character label on the open line,
the evaluation amounts to working out the action of the knot invariant
operator ${\mathcal I}_K: \sigma_1 \mapsto {\mathcal I}_K \sigma_1$.
Starting with $\sigma_1$, the final result of applying the nested
sequence of caps, braid crossings and then cups will give for
${\mathcal I}_K $ a product of the form
\begin{align}\label{eq:KnotOperatorIK}
   {\mathcal I}_K \sigma_1 
     &= \, \sum \sigma_{\kappa_1} \cdot a_1 \langle \sigma_{\kappa_2} 
        \cdot a_2|\sigma_2\rangle
        \langle \sigma_{\kappa_3} \cdot a_3|\sigma_3\rangle 
        \cdots \langle \sigma_{\kappa_m} \cdot a_m|\sigma_m\rangle.
\end{align}
Here $\kappa \in {\mathfrak S}_m$ is the image of $b_K$ under the projection
${\mathfrak B}_m \rightarrow  {\mathfrak S}_m$ reflecting the permutation
on strands induced by the braiding and the original labelling of strands
induced by the nested caps. In implementing the summations entailed in
${\texttt c}_\pi$, ${\texttt c}_\pi{}^{-1}$, there will be $\ell$ sets
of additional summations $\alpha_{(\pi)}{}^1, \alpha_{(\pi)}{}^2,
\cdots, \alpha_{(\pi)}{}^\ell $ symbolising the composite sums involved
in $r_\pi$; the summands appear repeated on correlated strands involved
in the various crossings. Moreover, for $w_-$ of these pairs,
\emph{one} of these occurrences will entail the antipode ${\sf S}(\cdot)$
reflecting the inverse crossings. The objects $a_i$ accompanying each
$\sigma_{\kappa i}$ thus represent the distribution of products of
these summands resulting from the application of the crossings.  

In the case of the braid representation of knot \texttt{8\_1} given
above, however, all crossings are positive and the braidings lead to
\begin{align}
\, \sigma_4\!\cdot\!a_1 
  &\otimes \sigma_5\!\cdot\!a_2 \otimes \sigma_1
   \!\cdot\!a_3 \otimes \sigma_2\!\cdot\!a_4 
   \otimes \sigma_3\!\cdot\!a_5 \otimes \sigma_2 
   \otimes \cdots \otimes \sigma_{5},
\end{align}  
with $a_1=\alpha_1\alpha_2 \alpha_7\alpha_9\alpha_{10}$, $a_2
=\alpha_1\alpha_2 \alpha_4\alpha_5\alpha_6\alpha_8$, $a_3
=\alpha_8\alpha_9$, $a_4=\alpha_6 \alpha_7$, and $a_5=\alpha_3\alpha_4$,
with the $\alpha \cdots \alpha $ standing for the
$\alpha_{(1)^{(\pi)}} \cdots \alpha_{(2)^{(\pi)}}$ summations over
Sweedler part plethysms associated with the 10 crossings in this case.
Including the bottom cups corresponding to the braid closures (on all
but the first lines) indeed leads to an expression of the form given
in (\ref{eq:KnotOperatorIK}) above, where the permutation
$\kappa \in {\mathfrak S}_5$ is the 5-cycle $\kappa = (4,5,1,2,3)$,
or $(42531)$ in cycle notation.

  We pursue the general analysis for the case that $\kappa$ is an
$m$-cycle, $\kappa = (1 \kappa_1 \kappa_2 \cdots \kappa_{m\!-\!1})$
in cycle notation. The strand $\sigma_1$ occurs in an outer product
with $a_{\kappa^{-1}1}$ where $\kappa^{-1}1=\kappa_{m\!-\!1}$; this
product is paired in a scalar product with the accompanying
$\sigma_{\kappa_{m\!-\!1}}$, which in turn occurs in a similar
arrangement with $a_{\kappa_{m\!-\!2}}$ and $\sigma_{\kappa_{m\!-\!2}}$.
Fixing on the sum over $\sigma_{\kappa_{m\!-\!1}}$ and denoting the
remaining entries by $\cdots$, we have
\begin{align}
  \sum \cdots \langle \sigma_1\cdot a_{\kappa_{m\!-\!1}} | 
  \sigma_{\kappa_{m\!-\!1}}\rangle \langle \sigma_{\kappa_{m\!-\!1}}
  \cdot a_{\kappa_{m\!-\!2}}|\sigma_{\kappa_{m\!-\!2}}\rangle  
 &\equiv \sum \cdots \langle \sigma_1\cdot 
   a_{\kappa_{m\!-\!1}}a_{\kappa_{m\!-\!2}}|\sigma_{\kappa_{m\!-\!2}}\rangle,
\end{align}
using of the resolution of the identity in the sum over $\sigma_{\kappa_{m\!-\!1}}$. Clearly this process can be iterated,  leaving the last sum over $\sigma_{\kappa_1} $ in the form
\begin{align}
\sum \sigma_{\kappa_1} \cdot 
   &a_1 \langle \sigma_1 \cdot  
    a_{\kappa_{m\!-\!1}}a_{\kappa_{m\!-\!2}}
    \cdots a_{\kappa_1}|\sigma_{\kappa_1} \rangle.
\end{align}
which yields
\begin{align}
{\mathcal I}_K \sigma_1 
   &= \sum \sigma_1 \cdot
      a_{\kappa_{m\!-\!1}}a_{\kappa_{m\!-\!2}} \cdots a_{\kappa_1},
\end{align}
and (the outer product being commutative) we recover the factor
$\prod_{i=1}^\ell a_i$. This contains the products of \emph{all} $\ell$
paired sets of $\alpha_{(\pi)}$ summands, including the terms which
carry an antipode associated with inverse crossings. Clearly then, as
the product is commutative, these factors can be rearranged as $w_+$
powers of $Q_\pi$, and $w_-$ powers of $Q_\pi{}^{-1}$, the result
being simply ${\mathcal I}_K = (Q_\pi)^w$ as a multiplicative factor.

If $\kappa$ is \emph{not} an $m$-cycle, then there are say $k>1$ sets
of strands (corresponding to the cycle structure of $\kappa$) which
do not communicate: the braid represents a link projection with two
or more $S^1$ components. In this case a similar argument to the
above goes through if the invariant is evaluated for a cut diagram
wherein one representative strand from \emph{each} cycle type is cut,
giving an link projection invariant operator ${\mathcal I}_L$ on
$\otimes^k \Lambda$. For example, if $k=2$ and the linked knots are
$K_1$ and $K_2$, the evaluation leads to
\begin{align}
  {\mathcal I}_L 
    &= (Q_\pi)^{w_1} \otimes (Q_\pi)^{w_2} \cdot (r_\pi){}^{w_{12}}
\end{align}
where $w_{12}$ is the \emph{linking number} of the two knots.

\vfill
\end{document}